\documentclass[
    aps,
    prl,
    twocolumn,
    10pt,
    a4paper,
    superscriptaddress,
    notitlepage,
    longbibliography,
    floatfix
]{revtex4-2}

\usepackage[english]{babel}
\usepackage{bm}
\usepackage{amsmath}
\usepackage{amssymb}
\usepackage{mathtools}
\usepackage{nicefrac}
\usepackage{xifthen}
\usepackage{etoolbox}
\usepackage{xstring}
\usepackage{textcomp}
\usepackage{upgreek}
\usepackage[pdftex]{graphicx}
\usepackage{wrapfig}
\usepackage{microtype}
\usepackage{enumitem}
\usepackage{xcolor}
\usepackage{gnuplottex}    
\usepackage{multirow}   
\usepackage{hyperref}
\usepackage{placeins}
\usepackage{csquotes}
\usepackage{physics}
\usepackage{siunitx}
\usepackage{booktabs}
\usepackage{txfonts}
\usepackage[capitalize]{cleveref}
\usepackage{orcidlink}
\usepackage{hyperref}
\hypersetup{colorlinks=true, citecolor=blue, urlcolor=blue, linkcolor=blue}

\newcommand{\kb}{k_\text{B}}

\newcommand{\iu}{\mathrm{i}}
\newcommand{\e}{\mathrm{e}}

\newcommand{\vect}[1]{\bm{{#1}}}

\newcommand{\adjVect}[1]{\adj{\vect{#1}}}

\newcommand{\trpVect}[1]{\trp{\vect{#1}}} 
\newcommand{\unitvect}[1]{\hat{\bm{{#1}}}}
\newcommand{\matr}[1]{\bm{{#1}}}

\DeclareMathOperator{\diag}{\mathop{\mathrm{diag}}}

\newcommand{\adj}[1]{#1^\dagger}

\newcommand{\inv}[1]{#1^{-1}}

\newcommand{\adjMatr}[1]{\adj{\matr{#1}}}
\newcommand{\trp}[1]{{#1}^\intercal} 
\newcommand{\trpMatr}[1]{\trp{\matr{#1}}} 
\newcommand{\conjMatr}[1]{\conj{\matr{#1}}}
\newcommand{\conj}[1]{#1^*}

\newcommand{\nns}[1]{\langle #1 \rangle} 

\newcommand{\timeReversal}{\mathcal{T}} 
\newcommand{\inversion}{\mathcal{P}} 
\newcommand{\translation}{\vect{\tau}} 
\newcommand{\spinSpaceTrafo}[3][]{
	[
		#2 \parallel #3
		\ifstrempty{#1}{}{\, | \, #1}
    ]
} 

\newcommand{\metric}{G} 
\newcommand{\parenthesisKron}[1]{\IfSubStr{#1}{+}{(#1)}{\IfSubStr{#1}{-}{(#1)}{#1}}}
\newcommand{\kron}[2]{
	\noexpandarg
	\delta_{
		\parenthesisKron{#1}
		\parenthesisKron{#2}
	}
} 

\newcommand{\hamil}{\mathcal{H}} 
\newcommand{\hmatr}{\matr{H}} 
\newcommand{\hMatr}{\hmatr} 

\newcommand{\nucs}{N_{\text{uc}}} 
\newcommand{\nbands}{N} 

\begin{document}
\title{Odd-Parity-Wave Magnons and Nonrelativistic Thermal Edelstein Effect}

\author{Robin R.~Neumann\orcidlink{0000-0002-9711-3479}}
\email[Correspondence email address: ]{robin.neumann@uni-muenster.de}
\affiliation{Institute of Solid State Theory, University of Münster, D-48149 Münster, Germany}
\affiliation{Institute of Physics and Halle-Berlin-Regensburg Cluster of Excellence CCE, Martin Luther University Halle-Wittenberg, D-06099 Halle (Saale), Germany}

\author{Rodrigo Jaeschke-Ubiergo\orcidlink{0000-0002-4821-8303}}
\affiliation{Institute of Physics, Johannes Gutenberg University Mainz, D-55128 Mainz, Germany}

\author{Ricardo Zarzuela\orcidlink{0000-0003-1765-1697}}
\affiliation{Institute of Physics, Johannes Gutenberg University Mainz, D-55128 Mainz, Germany}

\author{Libor Šmejkal\orcidlink{0000-0003-1193-1372}}
\affiliation{Max Planck Institute for the Physics of Complex Systems, 01187 Dresden, Germany}
\affiliation{Max Planck Institute for Chemical Physics of Solids, 01187 Dresden, Germany}

\author{Jairo Sinova\orcidlink{0000-0002-9490-2333}}
\affiliation{Institute of Physics, Johannes Gutenberg University Mainz, D-55128 Mainz, Germany}
\affiliation{Department of Physics, Texas A\&M University, College Station, TX, USA}

\author{Alexander Mook\orcidlink{0000-0002-8599-9209}}
\affiliation{Institute of Solid State Theory, University of Münster, D-48149 Münster, Germany}

\begin{abstract}
Odd-parity-wave magnets are noncollinear compensated magnets with spin-split band structure in the absence of spin-orbit coupling and dipolar interactions.
In contrast to altermagnets, their spin-polarized band structure breaks inversion symmetry, but preserves time-reversal symmetry rendering their spin texture odd in momentum space.
Here, we study the spin dynamics of the magnetic texture and compute the band structure and spin polarization of magnons.
We present minimal spin models of noncoplanar odd-parity-wave magnets purely stabilized by exchange interactions that host $p$- and $f$-wave spin textures for the magnetic excitations.
We demonstrate that two of these models exhibit collinear spin textures, i.e., the magnon spin polarization is restricted to a global (quantization) axis independent of the momentum giving rise to single-component odd-parity-wave magnetism, previously associated primarily with coplanar ground states.
Finally, the nonrelativistic magnonic thermal Edelstein effect—a nonequilibrium magnetization induced by a temperature gradient—is shown to exist for $p$-wave magnets in linear response and inherits its anisotropic angular dependence from the partial-wave character of the spin-polarized band structure.
Our findings suggest that insulating odd-parity-wave magnets are promising candidates for magnon spintronics applications.
\end{abstract}

\date{\today}
\maketitle

\paragraph{Introduction.}
The recent discovery of altermagnetism~%
\cite{%
    smejkal_beyond_2022,
    smejkal_emerging_2022%
},
and odd-parity-wave magnetism~%
\cite{%
    hellenes_unconventional_2024,%
    jungwirth_altermagnetism_2025%
},
has fundamentally reshaped our understanding of compensated magnetic systems.
These phases demonstrate that nontrivial spin-split electronic band structures can emerge \emph{without} spin-orbit coupling or net magnetization.
In altermagnets, the spin expectation value of an electron with momentum $\vect{k}$ in band $n$ is even under momentum inversion, satisfying $\vect{s}_{n \vect{k}} = \vect{s}_{n (-\vect{k})}$, whereas odd-parity-wave magnets host antisymmetric spin polarization, $\vect{s}_{n \vect{k}} = -\vect{s}_{n (-\vect{k})}$.
Such odd-parity-wave spin textures emerge without spin-orbit coupling when four key ingredients coincide: (i) broken combined inversion-time-reversal ($\inversion \timeReversal$) symmetry, (ii) broken inversion ($\inversion$) symmetry, (iii) a nonsymmorphic time-reversal symmetry ($\timeReversal \translation$) that pairs time reversal ($\timeReversal)$ with a fractional lattice translation ($\translation$)~%
\footnote{%
    In principle, the requirement of nonsymmorphic time-reversal symmetry can be relaxed by suppressing even-parity spin components via spin-group symmetries, and odd-parity-wave spin splitting can also arise in noncentrosymmetric materials~%
    \cite{%
        hayami_spontaneous_2020,
        hu_spin_2025,
        luo_spin_2025%
    }.
    Here, however, we focus exclusively on centrosymmetric odd parity with nonsymmorphic time-reversal symmetry, since in this setting the odd-wave spin splitting is generated purely by magnetic order. As a result, even in the presence of spin–orbit coupling, the odd-parity-wave character of the spin splitting is symmetry protected and remains intact as long as the magnetic order is preserved.%
},
and (iv) noncollinear magnetic order.
Collinear magnets feature either even-parity-wave spin splitting or spin-degenerate band structures owing to their spin-only symmetries~%
\cite{%
    smejkal_beyond_2022,%
    smejkal_emerging_2022%
}.
The momentum-dependent spin polarization and unconventional spin-charge interconversion of odd-parity-wave magnets have positioned them as promising candidates for next-generation spintronics~%
\cite{%
    brekke_minimal_2024,
    pari_nonrelativistic_2025,
    chakraborty_highly_2025,
    yu_odd-parity_2025%
}.
Yet one essential aspect remains unexplored: the \emph{dynamics} of these unconventional magnetic orders.

Magnetic dynamics in long-range ordered phases are governed by magnons -- the bosonic quasiparticles representing collective spin excitations.
They offer distinct advantages over electronic carriers for spin information transport, including low dissipation, long coherence, and full compatibility with insulating platforms \cite{Chumak2015}.
In conventional antiferromagnets, however, magnons of opposite spin polarization are degenerate and are easily mixed into spin-zero modes by weak spin–orbit coupling, limiting their usefulness for spin manipulation.
Altermagnets overcome this constraint by hosting exchange-driven even-parity-wave spin-split magnons whose spin polarization can survive the inclusion of spin-orbit coupling, enabling spin-selective and symmetry-controlled magnon transport~%
\cite{%
    Naka2019,
    smejkal_chiral_2023,
    Gohlke2023,
    cui_efficient_2023,
    weißenhofer_atomistic_2024,
    liu_chiral_2024,
    yu_chiral_2024,
    hoyer_spontaneous_2025,
    jost_chiral_2025,
    sourounis_efficient_2025,
    daghofer_altermagnetic_2025,
    lee_effective_2025,
    sarkar_spin-split_2025,
    siam_chiral-split_2025,
    khatua_magnon_2025,
    jin_interaction-driven_2025,
    yuan_quantum_2025,
    hoyer_altermagnetic_2025,
    syljuasen_quantum_2025,
    cichutek_spontaneous_2025,
    jin_strong_2025,
    eto_spontaneous_2025,
    beida_chiral_2025,
    kravchuk_chiral_2025,
    biniskos_systematic_2025,
    cichutek_quantum_2025,
    wiedmann_quantum_2025,
    yan_competitive_2025,
    bendin_d-wave_2025%
}.
The existence of altermagnetic magnons with even-parity-wave splitting naturally raises the question of whether odd-parity-wave magnetic order can imprint fundamentally new structures onto magnon excitations.

In this work, we close this gap by establishing the theoretical foundation of odd-parity-wave magnons.
We show that the odd-parity-wave character of the spin order is transferred to the magnon band structure, resulting in spin excitations with opposite spin polarization at $+\vect{k}$ and $-\vect{k}$.
This symmetry-protected locking of spin and propagation direction—normally associated with relativistic spin–orbit coupling in noncentrosymmetric systems~\cite{cheng_antiferromagnetic_2016,gitgeatpong_nonreciprocal_2017}—emerges here purely from exchange interactions.
Using minimal models based solely on bilinear and biquadratic isotropic Heisenberg exchange, we uncover $p$- and $f$-wave magnon spin textures with clear experimental signatures.
Most notably, we predict a nonrelativistic magnonic thermal Edelstein effect—a temperature-gradient-induced magnetization—that directly reflects the symmetry of the underlying magnon spin texture.

\begin{figure}
    \centering
    \includegraphics[width=\linewidth]{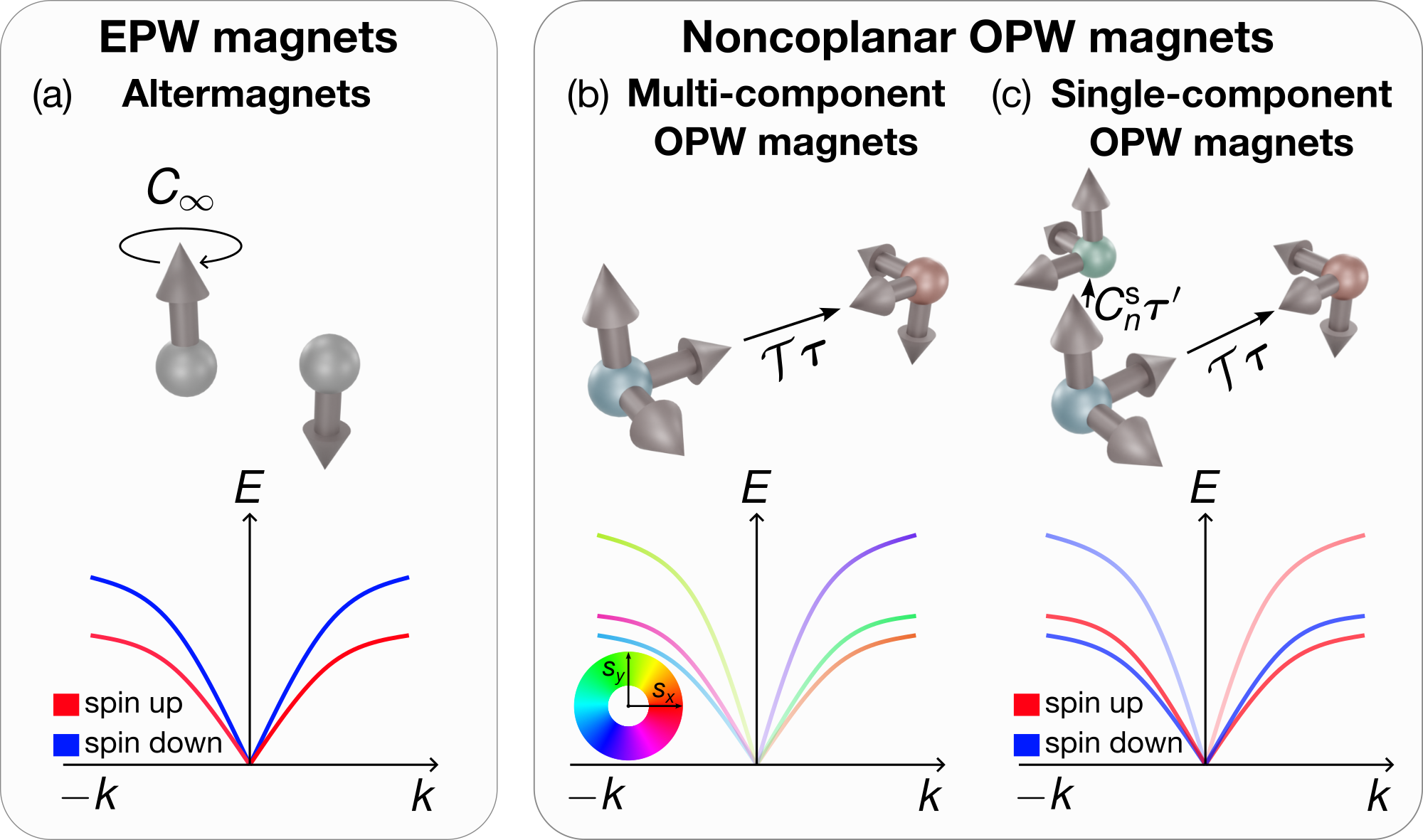}
    \caption{%
        Magnons in unconventional magnets.
        (a) Altermagnets possess collinear magnetic ground states giving rise to two linear Goldstone magnons with quantized spins and even-parity-wave (EPW) splitting.
        The spin is confined to a global quantization axis producing a collinear spin texture.
        (b),\,(c) Noncoplanar odd-parity-wave (OPW) magnets generally possess $\timeReversal \translation$ symmetry and exhibit three linear Goldstone magnons with nonquantized spins.
        (b) Multi-component odd-parity-wave magnets lack additional symmetries to constrain the magnon spin to a global axis.
        The magnon spin must be treated as a vector that lives in two- or three-dimensional space.
        (c) Single-component odd-parity-wave magnets are odd-parity-wave magnets with additional symmetries that constrain the magnon spin to a global axis.
        Although not quantized, the magnon spin forms a collinear spin texture and can be treated as a scalar.
        Originally proposed in coplanar systems~\cite{hellenes_unconventional_2024}, here we uncover single-component odd-parity-wave magnetism in noncoplanar systems, which exhibit a $n$-fold spin rotation $C_{n}^{\mathrm{s}}$ and a translation $\translation'$ with $n \geq 2$.
    }
    \label{fig:graphical_abstract}
\end{figure}
\paragraph{Magnons in unconventional magnets.}
Magnons in altermagnets [\cref{fig:graphical_abstract}(a)] owe their even-parity-wave spin polarization to the collinearity of the magnetic ground state.
The quantized magnon spin and inversion symmetry of the band structure follow from the collinear spin-only group~%
\footnote{Spin-only groups encompass those symmetry elements that only transform the spin space, but leave the real space invariant.}
$
    \mathbf{r}_{\rm so}
    =
    \mathrm{SO}(2)
    \times
    \mathbb{Z}_2^{
        \spinSpaceTrafo{C_2 \timeReversal}{\timeReversal}
    }
$,
where SO(2) contains all rotations of arbitrary angles about the collinear spin axis and
$
    \mathbb{Z}_2^{
        \spinSpaceTrafo{C_2 \timeReversal}{\timeReversal}
    }
    =
    \qty{
        \spinSpaceTrafo{E}{E},
        \spinSpaceTrafo{C_2 \timeReversal}{\timeReversal}
    }
$
contains the identity $E$ and $\spinSpaceTrafo{C_2 \timeReversal}{\timeReversal}$ that combines time reversal $\timeReversal$ with a two-fold rotation $C_2$ about an axis perpendicular to the collinear axis~\cite{smejkal_beyond_2022, smejkal_chiral_2023}.
In this framework of spin symmetries, the operations acting in spin space (left of \enquote{$\parallel$}) and those acting in real space (right of \enquote{$\parallel$}) are independent in the absence of spin-orbit coupling.

Magnons in odd-parity-wave magnets [\cref{fig:graphical_abstract}(b)] owe their antisymmetric spin polarization to a noncollinear spin arrangement in real space.
Noncollinear systems with the symmetry $\timeReversal \translation$ (short for $\spinSpaceTrafo[\translation]{\timeReversal}{\timeReversal}$, where the operation left of \enquote{$|$} is the point-group operation, while the operation right of \enquote{$|$} corresponds to a translation) are characterized by a spin translation group \cite{Watanabe2024}
\footnote{%
    Strictly speaking, $\mathbf{G}_{\rm so}$ is the point-group sector of the nontrivial spin translation group defined in Ref.~\cite{Watanabe2024}, which is obtained by projecting all translations into the identity.%
}
$
    \mathbf{G}_{\rm st}
    =
    {}^{\bar{1}}
    1
    =
    \qty{
        \spinSpaceTrafo{E}{E},
        \spinSpaceTrafo{\timeReversal}{\timeReversal}
    }
$.
This symmetry enforces an odd-parity-wave character in all spin components independently.
In contrast to altermagnets, odd-parity-wave magnets generally lack additional symmetries that restrict the dimension of the spin texture implying that the spin has to be treated as a vector rather than a scalar.
We call those odd-parity-wave magnets with a noncollinear spin texture in momentum space \enquote{multi-component odd-parity-wave magnets}.

Single-component odd-parity-wave magnetism, in addition to odd-parity-wave spin polarization, requires a collinear spin polarization in momentum space.
This was originally realized by considering a coplanar spin arrangement in real space~\cite{hellenes_unconventional_2024, jungwirth_altermagnetism_2025}, which leads to a spin-only group
$
    \mathbf{r}_{\rm so}
    =
    \mathbb{Z}_2^{
        \spinSpaceTrafo{C_{2} \timeReversal}{\timeReversal}
    }
$.
Here, $C_2$ is a two-fold rotation along the axis perpendicular to the coplanar spins.
This coplanar symmetry, together with $\timeReversal \translation$, enforces a collinear spin texture in momentum space whose polarization is perpendicular to the coplanar spins.

In this Letter, we propose an alternative mechanism to realize single-component odd-parity-wave magnetism in \emph{noncoplanar} magnets [\cref{fig:graphical_abstract}(c)].
Here, the noncoplanar spins are engineered to have a high-order spin-translation group
$
    \mathbf{G}_{\rm st}
    =
    {}^{n} 1
    {}^{\bar{1}} 1
$,
generated by the symmetries
$
    \timeReversal \translation
$
and
$
    C_{n}^{\mathrm{s}} \translation'
    =
    \spinSpaceTrafo[\translation']{C_{n}}{E}
$.
This arrangement generates collinear spin polarization of the magnon bands along the spin rotation axis of $C_{n}$ ($n \geq 2$), despite the noncoplanarity in real-space.
In the following we introduce models for general odd-parity-wave magnons hosted by noncoplanar magnetic ground states in three scenarios: (i) multi-component odd-parity $p$-wave magnet with noncollinear spin texture in momentum space; (ii) single-component odd-parity $p$-wave magnet with collinear spin texture in momentum space; (iii) single-component odd-parity $f$-wave magnet with collinear spin texture in momentum space.

\paragraph{Model of multi-component $p$-wave magnet.}
\begin{figure*}
    \includegraphics[width=\linewidth]{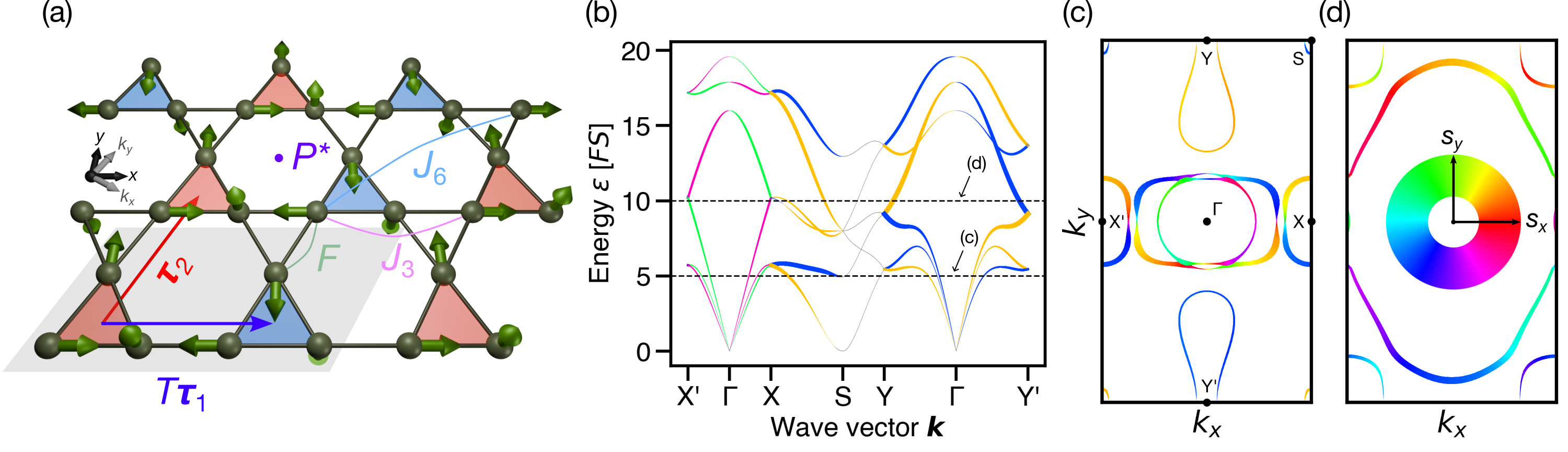}
    \caption{%
        Multi-component odd-parity $p$-wave kagome model.
        (a) Lattice structure, magnetic ground state configuration, and magnetic interactions $F$, $J_3$, $J_6$.
        The gray transparent quadrangle indicates the magnetic unit cell and the red and blue triangles distinguish the opposite local spin configurations related by time reversal.
        (b) Magnon band structure along a high-symmetry path in the first Brillouin zone [high-symmetry points indicated in (c)].
        (c),~(d) Isoenergy lines for (c) $\varepsilon = 5 F S$ and (d) $\varepsilon = 10 F S$.
        In panels (b)--(d), the color represents the spin angle within the $xy$ plane [see inset of (d)], while the line thickness corresponds to the magnitude of the spin.
        Note that the $k_x$ and $k_y$ axes are rotated by \SI{30}{\degree} with respect to the $x$ and $y$ axes [see panel~(a)].
        The parameters are $J_3 = 2 J_6 = F$.
    }
    \label{fig:noncoplanar_kagome}
\end{figure*}

First, we consider the noncoplanar magnetic texture on the kagome lattice displayed in \cref{fig:noncoplanar_kagome}(a)~\cite{hellenes_unconventional_2024}.
The spins in each triangular plaquette are mutually orthogonal.
This local magnetic structure is repeated along the $\translation_2$ direction (ferroic ordering), while it is staggered along the $\translation_1$ direction (antiferroic ordering), doubling the unit cell of the kagome lattice and installing time-reversal symmetry within the point group.
Importantly, the order breaks inversion symmetry because a point inversion at the center of the hexagon flips the in-plane spin components, but leaves the out-of-plane component invariant.
This operation can be remedied by a spin rotation about the $z$ axis by \SI{180}{\degree}, which makes $\inversion^* = \spinSpaceTrafo{C_{2z}}{\inversion}$ a symmetry of the system.
$\inversion^*$ maps $\vect{k}$ to $-\vect{k}$, but leaves $s_z$ invariant such that $s_z$ is even in momentum space and, due to the additional $\timeReversal \translation_1$ symmetry, $s_z$ vanishes (or must be degenerate).
Thus, this magnetic texture is noncoplanar (i.e., three-dimensional) in real space but causes a coplanar (i.e., two-dimensional) quasiparticle spin texture in momentum space.
Overall, the spin point group is ${}^{\bar{1}}1{}^{2_{x}}m{}^{2_y}m{}^1m$ [see Supplemental Material (SM)~\cite{supplement}].

To investigate magnons in this system, the described magnetic configuration needs to be realized as the classical ground state of a spin Hamiltonian, which we construct as
\begin{align}
    \hamil
    &=
    \frac{F}{2 S^2 \hbar^4}
    \sum_{\nns{ij}_1}
    \qty(
        \vect{S}_i
        \cdot
        \vect{S}_j
    )^2
    +
    \frac{J_3}{2 \hbar^2}
    \sum_{\nns{ij}_3}
    \vect{S}_i
    \cdot
    \vect{S}_j
    +
    \frac{J_6}{2 \hbar^2}
    \sum_{\nns{ij}_6}
    \vect{S}_i
    \cdot
    \vect{S}_j
    ,
    \label{eq:spin_hamil_kagome}
\end{align}
where $\hbar$ is the reduced Planck constant, $\vect{S}_i$ is the spin operator with spin quantum number $S$ on site $i$, and the sums run over first ($\langle ij \rangle_1$), third ($\langle ij \rangle_3$), and sixth ($\langle ij \rangle_6$) nearest neighbors.
The biquadratic interaction $F > 0$ stabilizes the classical \SI{90}{\degree} order between nearest neighbors. The bilinear Heisenberg interactions $J_3$ and $J_6$ form a network of three disconnected triangular lattices, where they couple collinear spins in \cref{fig:noncoplanar_kagome}(a), which are parallel and antiparallel in a stripe order on each of the three triangular lattices. The ordering vector $\vect{Q}$ of the sublattice stripes is identical for the three sublattices.
Classically, the stripe order is stable in the regime where $J_3 > 0$ and $J_3 / 8 < J_6 < J_3$~%
\cite{%
    jolicoeur_ground-state_1990,
    chubukov_order_1992,
    zhu_disorder-induced_2017%
}, although quantum fluctuations generally shift the phase boundaries~%
\cite{%
    zhu_spin_2015,
    zhu_disorder-induced_2017,
    gallegos_phase_2025%
}.
In the following, we work with the parameters $J_3 = 2 J_6 = F$, where the stripe order is stable against quantum fluctuations.
For details of the magnon dispersion and spin polarization calculations within a large-$S$ spin-wave theory, please refer to the End Matter and the SM~\cite{supplement}.

The magnon band structure and magnon spin polarization are shown in \cref{fig:noncoplanar_kagome}(b) along a high-symmetry path in the first Brillouin zone [high-symmetry points indicated in \cref{fig:noncoplanar_kagome}(c)].
Since the magnetic unit cell contains six spins [indicated by the gray quadrangle in \cref{fig:noncoplanar_kagome}(a)], there are six magnon bands, three of which are linear Goldstone modes as expected for a magnetically compensated ground state that spontaneously breaks all three generators of the SO(3) symmetry of $\hamil$~%
\cite{%
    goldstone_broken_1962,
    nielsen_how_1976,
    watanabe_unified_2012%
}.
Moreover, we observe an \emph{accidental} zero-energy mode at the S point not protected by the Goldstone theorem, which is known for stripe-ordered magnets on the triangular lattice and which is gapped out by quantum fluctuations~%
\cite{%
    chubukov_order_1992,
    willsher_magnetic_2023%
}.

The in-plane orientation of the spin is encoded in the color of the bands, while the line thickness represents the magnitude of the spin polarization.
Indeed, the magnon spin polarization is odd in momentum space, i.e., $\vect{s}_{n \vect{k}} = -\vect{s}_{n (-\vect{k})}$ for each band $n$ and wave vector $\vect{k}$.
This is reflected in the complementary colors at opposite momenta in \cref{fig:noncoplanar_kagome}(b).
Two isoenergy cuts at $\varepsilon = 5 F S$ and $\varepsilon = 10 F S$ are shown in \cref{fig:noncoplanar_kagome}(c) and (d), respectively.
Tracing a closed isoenergy line, the spin polarization winds exactly once in the $xy$ plane, demonstrating the $p$-wave nature of the spin texture.
Notably, the spin polarization is not restricted to a global axis, i.e., there is no collinear spin texture in this model.
Since even for a closed isoenergy line the magnitude of the spin does not vanish, the nodal lines depend on the spin direction as well as the energy and are generally curved.
This is because there is no symmetry fixing the nodal line.

\begin{figure*}
    \includegraphics[width=\linewidth]{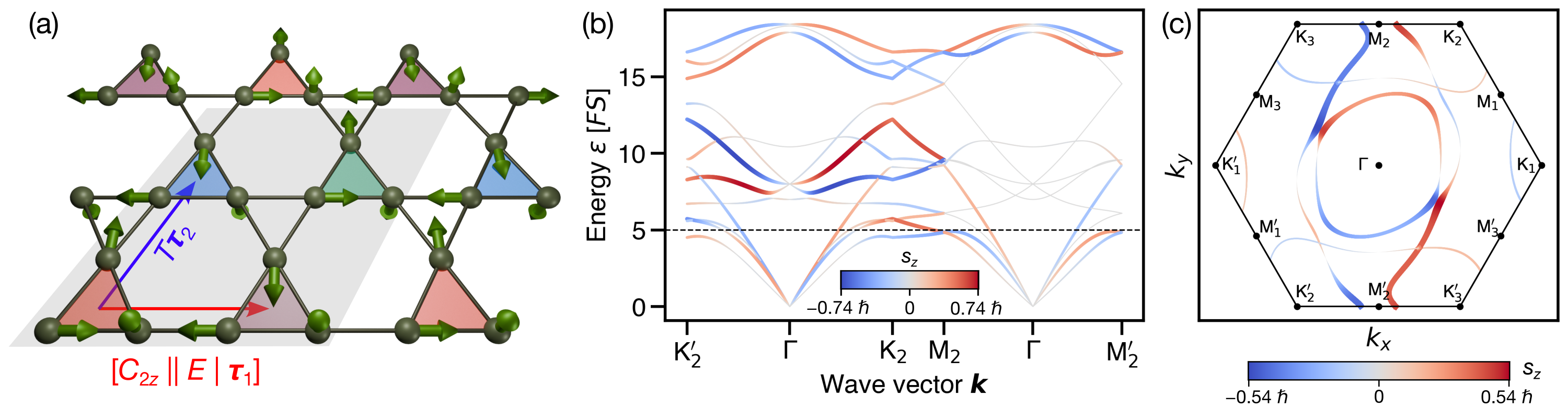}
    \caption{%
        Single-component odd-parity $p$-wave kagome model.
        (a) Lattice structure and magnetic ground state configuration.
        The gray transparent quadrangle indicates the magnetic unit cell and the colored triangles distinguish the 4 different local spin configurations related by (the combination of) time reversal and 2-fold spin rotation.
        (b) Magnon band structure along a high-symmetry path in the first Brillouin zone [high-symmetry points indicated in (c)].
        (c) Isoenergy lines for $\varepsilon = 5 F S$ in the first Brillouin zone (hexagon).
        In panels~(b) and (c), the color represents the $z$ component of the spin (see color bars), while the line thickness corresponds to its magnitude.
        The parameters are $J_3 = 2 J_6 = F$.
    }
    \label{fig:oop_kagome}
\end{figure*}

\paragraph{Model of single-component $p$-wave magnet.}
Next, we consider a distinct classical ground state.
We abandon the ferroic ordering along the $\translation_2$ direction and instead stagger (i) the in-plane spins along $\translation_1$ and (ii) all spin components along $\translation_2$, as visualized in \cref{fig:oop_kagome}(a).
This magnetic configuration constitutes a classical ground state of the Hamiltonian $\hamil$ in \cref{eq:spin_hamil_kagome} and is degenerate with the previously discussed state, since the spins remain mutually orthogonal and the collinear sublattices continue to form a stripe pattern. The key difference is that the $\vect{Q}$ vectors associated with the three orthogonal stripe orders forming the kagome lattice are no longer identical. Because the relative orientations of these $\vect{Q}$ vectors do not influence the ground-state energy—spins belonging to different stripe patterns are coupled only via biquadratic interactions—we employ the same model parameters as before.
The aforementioned staggering introduces $\spinSpaceTrafo[\translation_1]{C_{2z}}{E}$ as a spin space group symmetry, enforcing the vanishing of the \emph{in-plane} spin components of the magnons, while breaking the $P^*$ symmetry that previously suppressed $s_z$.
Thus, the spin point group  ${}^{2_z}1 {}^{\bar{1}}1 {}^{2_y}2/{}^{1}m$ (see SM~\cite{supplement}) enforces an exclusive out-of-plane spin polarized magnon band structure with odd parity, i.e., $s_{n \vect{k}}^z = -s_{n (-\vect{k})}^z$.

This is confirmed by the linear spin-wave calculation shown in \cref{fig:oop_kagome}(b).
We identify 12 magnon bands and four zero-energy modes among them, of which three are enforced by the Goldstone theorem—the remaining fourth mode is expected to be accidental and to get gapped by fluctuations.
The bands are colored according to their out-of-plane spin component, while the line thickness again represents the magnitude of the spin polarization.
We find no nodal lines among the high-symmetry directions, where the spin polarization of all bands vanishes, but several bands are nearly unpolarized between $\overline{\mathrm{M}_2 \Gamma \mathrm{M}_2'}$ and all bands are unpolarized at $\Gamma$.
Note that the Brillouin zone has hexagonal shape, but the 3-fold rotational symmetry is broken, giving rise to inequivalent high-symmetry paths connecting $\Gamma$ and the K and M points indicated in \cref{fig:oop_kagome}(c)~%
\footnote{%
    In principle, it is sufficient to distinguish between K and K' as well as M and M' because of translational invariance.
    For example, K$_1$ and K$_3$ are related by a reciprocal lattice vector.
    However, the high-symmetry lines $\overline{\mathrm{\Gamma K_1}}$ and $\overline{\mathrm{\Gamma K_3}}$ are not equivalent, which is why we distinguish between equivalent points.%
}.

This symmetry breaking is further reflected in the isoenergy cut at $\varepsilon = 5 F S$ shown in \cref{fig:oop_kagome}(c).
One verifies the $p$-wave character of the energy by the opposite spin polarization at opposite momenta for all bands.
For the innermost closed isoenergy line, the spin changes sign twice—representing a single nodal line.
For other bands, however, the spin polarization may change sign multiple times accidentally.
These additional zeros are not protected by symmetry and may be removed by perturbations of the Hamiltonian.
Only the spin degeneracy at the \enquote{nodal points} $\Gamma$, M$_1$, M$_2$, and M$_3$ are protected by time-reversal symmetry. 
We emphasize that while this system is noncoplanar, it is an single-component odd-parity-wave magnet due to the collinearity of the spin polarization in momentum space, that is a consequence of its rich spin translation group ${}^{2_z}1 {}^{\bar{1}}1$.

\paragraph{Model of single-component $f$-wave magnet.}
\begin{figure*}
    \includegraphics[width=\linewidth]{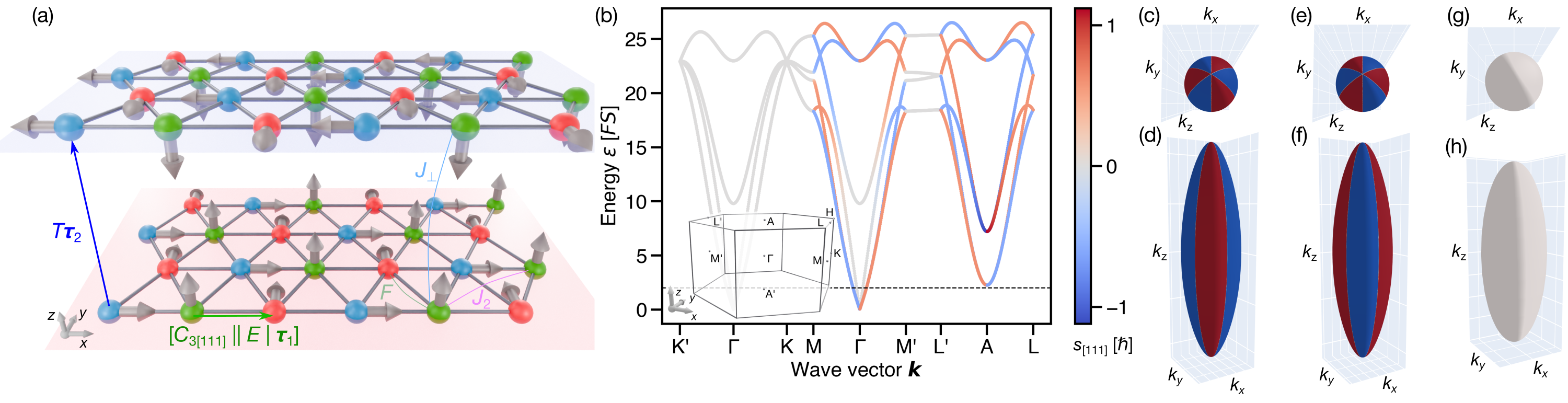}
    \caption{
        Three-dimensional $f$-wave model consisting of stacked triangular lattices.
        (a) Real-space lattice, magnetic ground state configuration, and magnetic interactions $F$, $J_2$, $J_\perp$.
        Collinear magnetic sites are colored alike.
        The two layers in the magnetic unit cell are related by time reversal.
        (b) Band structure along a high-symmetry path in the first Brillouin zone (see inset).
        (c)--(h) Isoenergy surfaces for $\varepsilon = 2 F S$ [dashed line in panel~(b)].
        Panels (c),~(d) show band 1, (e),~(f) show band 2, and (g),~(h) show band 3 starting from the lowest energy in panel~(b).
        The color in panels (b)--(h) represents the [111] component of the spin (see color bar).
        The parameters are $J_2 = -J_\perp = -F$.
    }
    \label{fig:ortho_triangular}
\end{figure*}

Although we have previously observed magnon bands with more than one nodal line, these zeros were accidental and, thus, not considered higher-order-wave magnets.
To realize a symmetry-protected $f$-wave magnon spin texture, we employ 3-fold rotation symmetry that triples the number of nodal lines/planes.
For this purpose, we consider a three-dimensional lattice composed of stacked triangular layers as shown in \cref{fig:ortho_triangular}(a).
The magnetic unit cell contains three orthogonal spins per layer and two layers related by time reversal.
Additionally, rotating the spins about the [111] axis by \SI{120}{\degree} interchanges the three spins within each layer, which can be compensated by a translation $\translation_1$.
This $\spinSpaceTrafo[\translation_1]{C_{3[111]}}{E}$ symmetry enforces a collinear reciprocal-space spin texture along the [111] spin-space direction.
Furthermore, there are three nodal planes at \SI{30}{\degree}, \SI{150}{\degree}, and \SI{270}{\degree} with respect to the $x$ axis that include the $z$ axis and are related by 3-fold rotation symmetry.
Overall, the spin point group is $^{\bar{3}} 1 ^m 6 / ^1 m ^m m ^1 m$ (see SM~\cite{supplement}).

To stabilize the magnetic texture without frustration, we consider biquadratic nearest-neighbor coupling $F > 0$ and ferromagnetic next-nearest-neighbor coupling $J_2 < 0$ within each layer, as well as antiferromagnetic interlayer coupling $J_\perp > 0$ between nearest neighbors:
\begin{align}
    \hamil
    &=
    \frac{F}{2 S^2 \hbar^4}
    \sum_{\langle ij \rangle_1}
    \qty(
        \vect{S}_i
        \cdot
        \vect{S}_j
    )^2
    +
    \frac{J_2}{2 \hbar^2}
    \sum_{\langle ij \rangle_2}
    \vect{S}_i
    \cdot
    \vect{S}_j
    +
    \frac{J_\perp}{2 \hbar^2}
    \sum_{\langle ij \rangle_\perp}
    \vect{S}_i
    \cdot
    \vect{S}_j
    .
\end{align}
We choose the parameters as $J_2 = -J_\perp = -F$.

The angles of the nodal planes coincide with the $\overline{\Gamma \mathrm{K}}$ and $\overline{\Gamma \mathrm{K}'}$ directions, rendering the 6 magnon bands unpolarized along these lines [cf.~\cref{fig:ortho_triangular}(b)].
At the nodal surfaces, the lower two and the upper two bands become degenerate, while the middle two bands remain isolated, but their spin expectation values vanish.

Three bands intersect with the energy level $\varepsilon = 2 F S$ indicated by the dashed line in \cref{fig:ortho_triangular}(b).
Their isoenergy surfaces are shown in \cref{fig:ortho_triangular}(c)--(h).
Two of them are fully spin polarized and touch at the nodal planes, while one is isolated from the others and has a weak spin polarization because the isoenergy surface is closely surrounding the $\Gamma$ point where its spin exactly vanishes.
To summarize, this system corresponds to an single-component $f$-wave magnet, due to the \textit{collinearity} of the spin polarization in momentum space,  which emerges from the noncoplanar real-space spin configuration due to the higher-order spin translation group ${}^{\bar{3}}1$ (see SM~\cite{supplement}), rather than from the coplanar spin-only group as in Ref.~\cite{jungwirth_altermagnetism_2025}.

\paragraph{Nonrelativistic magnonic Edelstein effect.}
Since the magnon band structure is spin polarized, an asymmetric population of the magnon states can induce a finite net magnetization.
As an example, the application of a temperature gradient can modify the magnon distribution and thereby lift the magnetic compensation generating a nonequilibrium magnetization according to
$
    \expval{S_{\mu}}
    =
    \sum_{\nu}
    \chi_{\mu \nu}
    (-\nabla_\nu T)
$,
which is known as the thermal Edelstein effect.

\begin{figure}
    \centering
    \includegraphics[width=\linewidth]{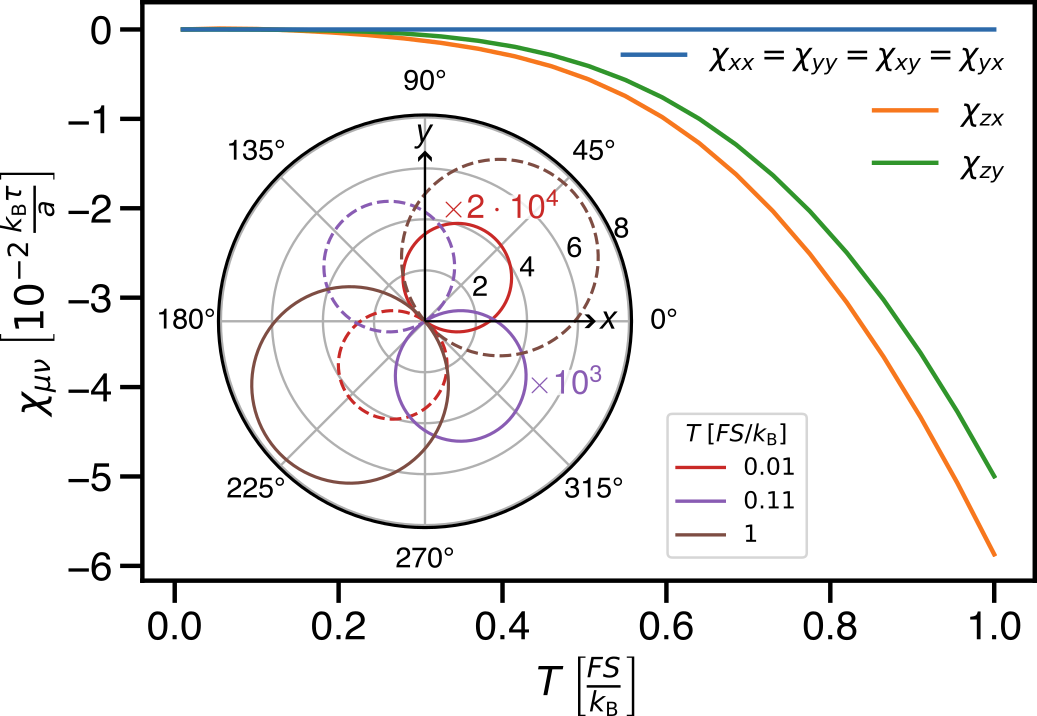}
    \caption{%
        Temperature-dependent linear thermal Edelstein effect of the single-component $p$-wave magnet with parameters $J_3 = 2 J_6 = F$.
        Here, $\tau$ is the relaxation time and $a$ is the nearest-neighbor distance.
        Inset: Nonequilibrium $z$ spin polarization as a function of the direction of $-\grad T$.
        The angle of \SI{0}{\degree} corresponds to $-\grad T \parallel \unitvect{x}$.
        Different colors correspond to different temperatures (see legend).
        For better visibility, the data for $T = \num{0.01}$ (red curve) and $\num{0.11}\,F S / \kb$ (purple curve) have been rescaled by a constant factor (see annotations).
        Solid and dashed lines indicate positive and negative sign, respectively.%
    }
    \label{fig:edelstein}
\end{figure}
We have computed the linear-response coefficients $\chi_{\mu\nu}$ in \cref{fig:edelstein} for the single-component $p$-wave model [recall Fig.~\ref{fig:oop_kagome}].
Because the magnon spin is fully oriented along the $z$ axis, there are only two nonzero components ($\chi_{zx}$ and $\chi_{zy}$), which are activated by temperature~%
\footnote{%
    Although the magnetic order is not stable at nonzero temperatures according to the Mermin-Wagner theorem~\cite{mermin_absence_1966}, a ferroic stacking with trivial interlayer coupling would stabilize the magnetic order without qualitatively changing the Edelstein response.%
}.
In the inset of \cref{fig:edelstein}, we present the dependence of the nonequilibrium spin polarization on the direction of the applied temperature gradient.
For a given temperature, we respectively find one lobe with positive (solid line) and one lobe with negative sign (dashed line) reminiscent of a $p$ orbital.
This shape implies a prominent anisotropy with large responses for temperature gradients applied in the direction of the lobes.
Upon ramping up the temperature, the $p$-orbital-like pattern rotates by approximately \SI{180}{\degree} and expands in size.
The elements with $\mu = x, y$ vanish because the magnon bands are unpolarized with respect to $s_x$ and $s_y$ rendering the corresponding Edelstein response zero.
This situation is reversed in the multi-component $p$-wave magnet without the collinear spin texture, where only the $\mu = x, y$ components survive (see SM~\cite{supplement}).

\paragraph{Discussion and conclusion.}
We have presented minimal models with ground states that support odd-parity-wave magnetism relying only on bilinear Heisenberg and biquadratic interactions.
The kagome models realize $p$-wave magnon spin polarization, while the stacked triangular lattice model hosts an $f$-wave spin texture.
Notably, two of these \emph{noncoplanar} models exhibit collinear spin textures~%
\cite{%
    bernevig_exact_2006,
    schliemann_colloquium_2017,
    tao_persistent_2018,
    tao_perspectives_2021,
    tenzin_persistent_2025%
},
i.e., the magnon spin polarization is restricted to a global axis independent of the momentum.
Hence, these noncoplanar states realize single-component odd-parity-wave magnetism previously associated primarily with coplanar ground states~\cite{hellenes_unconventional_2024}, demonstrating the possibility of engineering reciprocal-space spin textures by careful design of the magnetic ground state with symmetries that are not captured by the spin-only group.
This allows control over the nonrelativistic thermal Edelstein effect, which directly reflects the symmetries of the spin texture.
The magnonic Edelstein effect thus provides both a diagnostic tool for identifying $p$-wave magnetism in insulators and a potential mechanism for energy-efficient spin control in insulating systems.
Our findings enlarge the set of potential material candidates for single-component odd-parity-wave magnets two-fold by extending it not only to insulating magnets, but also to noncoplanar magnets.

Introducing magnons in odd-parity-wave magnets opens up several interesting questions for future research.
The models presented herein have been idealized for simplicity, but may serve to predict unique signatures of odd-parity-wave dynamics in inelastic neutron spectroscopy/polarimetry~%
\cite{%
    brockhouse_scattering_1957,
    saenz_spin_1962,
    mcclarty_observing_2025,
    faure_altermagnetism_2025%
},
neutron spin echo spectroscopy~%
\cite{%
    mezei_neutron_1972,
    hicks_magnetic_2019,
    keller_neutron_2021%
},
magneto-Raman spectroscopy~%
\cite{%
    fleury_scattering_1968,
    cottam_theory_1972,
    sugai_two-magnon_1988%
},
and in spin-polarized electron energy loss spectroscopy~%
\cite{%
    plihal_spin_1999,
    ibach_novel_2003,
    vollmer_spin-polarized_2003,
    vollmer_spin-polarized_2004%
}.
As a next step, one may get closer to realistic materials by including spin-orbit coupling, which can be essential to realize the noncollinear magnetic ground states and to identify insulating candidates, e.g., by using ab-initio methods.

By establishing that odd-parity-wave magnets can host spin-polarized magnons with protected antisymmetric textures, this work opens a new pathway for magnon spintronics that combines the advantages of insulating materials with the symmetry-protected odd-parity-wave spin-momentum locking previously exclusive to electronic systems. Our findings suggest that the dynamics of odd-parity-wave magnets represents a practical platform for low-dissipation spin information processing, with potential applications in thermal spin logic and quantum information technologies.

\begin{acknowledgments}
\paragraph{Acknowledgments.}
This work was funded by the German Research Foundation (DFG) as part of the German Excellence Strategy EXC3112/1-533767171 (Center for Chiral Electronics) and through TRR 277-328545488 (Project No. B04), TRR 173-268565370 (Projects No. A03 and B13), TRR 288-422213477 (Projects No. A09 and B05), and Project No.~504261060 (Emmy Noether Programme). We acknowledge support by the Dynamics and Topology Center (TopDyn) funded by the State of Rhineland-Palatinate.
\end{acknowledgments}

\paragraph{Data availability.}
Upon reasonable request, the data associated with this article can be made available on Zenodo~\cite{zenodo}.

\bibliography{refs.bib}

\appendix
\section*{End Matter}
\subsection{Holstein-Primakoff transformation}
Here, we present the general concept of linear spin-wave theory and refer to the Supplemental Material~\cite{supplement} for the expressions for the specific models presented in the main text.
The starting point is the classical ground state given by the orientations $\unitvect{z}_i$ of the local spins at sites $i$.
Using the Holstein-Primakoff transformation~\cite{holstein_field_1940}, the spin operators are expanded as~\cite{toth_linear_2015,rau_pseudo-goldstone_2018}
\begin{align}
    \frac{\vect{S}_i}{\hbar}
    &=
    \qty(
        S_i
        -
        \adj{a}_i
        a_i
    )
    \unitvect{z}_i
    +
    \sqrt{S_i}
    \qty[
        \adj{a}_i
        \qty(
            1
            -
            \frac{n_i}{2 S_i}
        )^{\nicefrac{1}{2}}
        \unitvect{e}_i^+
        +
        \qty(
            1
            -
            \frac{n_i}{2 S_i}
        )^{\nicefrac{1}{2}}
        a_i
        \unitvect{e}_i^-
    ]
    \label{eq:hpt}
\end{align}
where $S_i$ is the spin quantum number, $\hbar$ the reduced Planck constant, $a_i$ and $\adj{a}_i$ bosonic annihilation and creation operators, $n_i = \adj{a}_i a_i$ the number operator and $\unitvect{e}_i^\pm = (\unitvect{x}_i \pm \iu \unitvect{y}_i) / \sqrt{2}$ auxiliary complex unit vectors where $\unitvect{x}_i$, $\unitvect{y}_i$, $\unitvect{z}_i$ form a (right-handed) tripod.

Because of translation symmetry, we can label each site $i$ by the unit cell $u$ and the sublattice $m$ and it is useful to describe the system in Fourier space:
\begin{subequations}
\begin{align}
    a_{m \vect{k}}
    &=
    \frac{1}{\sqrt{\nucs}}
    \sum_{u}
    \e^{-\iu \vect{k} \cdot (\vect{R}_u + \vect{r}_m)}
    a_{m u}
    ,
    \\
    \adj{a}_{m \vect{k}}
    &=
    \frac{1}{\sqrt{\nucs}}
    \sum_{u}
    \e^{\iu \vect{k} \cdot (\vect{R}_u + \vect{r}_m)}
    \adj{a}_{m u}
    .
\end{align}
\end{subequations}
Here, $\vect{R}_u$ is the position of the $u$-th unit cell and $\vect{r}_m$ is the position of the $m$-th sublattice within a unit cell.
In this representation, the bilinear Hamiltonian is diagonal in $\vect{k}$ and can be written as
\begin{align}
    \hamil^{(2)}
    &=
    \frac{1}{2}
    \sum_{\vect{k}}
    \adjVect{\phi}_{\vect{k}}
    \hMatr(\vect{k})
    \vect{\phi}_{\vect{k}}
    ,
    \label{eq:bilinear_hamil}
\end{align}
where
\begin{align}
    \adjVect{\phi}_{\vect{k}}
    =
    \mqty(
        \adj{a}_{1 \vect{k}} & \dots & \adj{a}_{\nbands \vect{k}}
        &
        a_{1 (-\vect{k})} & \dots & a_{\nbands (-\vect{k})}
    )
\end{align}
is the Nambu spinor with $2 \nbands$ entries ($\nbands$ is the number of sublattices) and
\begin{align}
    \hMatr(\vect{k})
    =
    \begin{pmatrix}
        \matr{A}(\vect{k}) & \matr{B}(\vect{k})
        \\
        \conjMatr{B}(-\vect{k}) & \conjMatr{A}(-\vect{k})
    \end{pmatrix}
    \label{eq:bilinear_kernel}
\end{align}
is the Hamiltonian's kernel.
Its structure is in principle arbitrary due to the redundancy of the Nambu space description, but one commonly chooses the one in \cref{eq:bilinear_kernel} to install particle-hole symmetry.
The Hermiticity of the Hamiltonian implies $\matr{A}(\vect{k}) = \adjMatr{A}(\vect{k})$ and $\matr{B}(\vect{k}) = \trpMatr{B}(-\vect{k})$.
The concrete expressions of $\matr{A}(\vect{k})$ and $\matr{B}(\vect{k})$ depend on the interactions embedded in $\hamil$ and the magnetic ground state.
Below, we provide general expressions for bilinear and biquadratic couplings.

In general, additive constants appear in \cref{eq:bilinear_hamil}, which correct the classical ground-state energy.
Since we are interested in the magnon band structure, we omit those constants herein.

\subsection{Bogoliubov transformation}
As a next step, the diagonalization of the kernel yields the one-particle energies:
\begin{align}
    \begin{split}
    \adjMatr{T}(\vect{k})
    \hMatr(\vect{k})
    \matr{T}(\vect{k})
    &=
    \matr{E}(\vect{k})
    \\
    &=
    \diag
    \mqty(
        \varepsilon_{1 \vect{k}} & \dots & \varepsilon_{\nbands \vect{k}}
        &
        \varepsilon_{1 (-\vect{k})} & \dots & \varepsilon_{\nbands (-\vect{k})}
    )
    ,
    \end{split}
\end{align}
such that new normal-mode Nambu spinors can be defined:
\begin{align}
    \adjVect{\psi}_{\vect{k}}
    =
    \adjVect{\phi}_{\vect{k}}
    \inv{\qty[
        \adjMatr{T}(\vect{k})
    ]}
    =
    \mqty(
        \adj{b}_{1 \vect{k}} & \dots & \adj{b}_{\nbands \vect{k}}
        &
        b_{1 (-\vect{k})} & \dots & b_{\nbands (-\vect{k})}
    )
    .
\end{align}
Since the diagonalization must leave the bosonic commutation rules intact, $\matr{T}(\vect{k})$ must be paraunitary, i.e.,~%
\cite{%
    colpa_diagonalization_1978,
    shindou_topological_2013%
}%
\begin{align}
    \adjMatr{T}(\vect{k})
    \matr{\metric}
    \matr{T}(\vect{k})
    =
    \matr{\metric}
    ,
\end{align}
where
\begin{align}
    \matr{\metric}
    =
    \diag
    \mqty(
    1 & \dots & 1 & -1 & \dots & -1
    )
    .
\end{align}

\subsection{Bilinear interactions}
Considering generic bilinear spin-spin interactions described by
\begin{align}
    \hamil
    =
    \frac{1}{2 \hbar^2}
    \sum_{m n}
    \sum_{u v}
    \trpVect{S}_{m u}
    \matr{J}_{m n}(\vect{R}_m - \vect{R}_n)
    \vect{S}_{n v}
    ,
\end{align}
where the sums run over all pairs of sublattices $m$ and $n$ and unit cells $u$ and $v$, the bilinear magnon Hamiltonian amounts to~\cite{toth_linear_2015,rau_pseudo-goldstone_2018}
\begin{subequations}
\begin{align}
    A_{m n}(\vect{k})
    &=
    \sqrt{S_m S_n}
    \trp{\qty[\unitvect{e}_m^+]}
    \matr{\mathcal{J}}_{mn}(\vect{k})
    \unitvect{e}_n^-
    -
    \kron{m}{n}
    \sum_{l = 1}^{\nbands}
    S_l
    \trp{\unitvect{z}}_m
    \matr{\mathcal{J}}_{ml}(\vect{0})
    \unitvect{z}_l
    ,
    \\
    B_{m n}(\vect{k})
    &=
    \sqrt{S_m S_n}
    \trp{\qty[\unitvect{e}_m^+]}
    \matr{\mathcal{J}}_{mn}(\vect{k})
    \unitvect{e}_n^+
    ,
\end{align}
\end{subequations}
where
\begin{align}
    \matr{\mathcal{J}}_{m n}(\vect{k})
    =
    \e^{\iu \vect{k} \cdot (\vect{r}_m - \vect{r}_n)}
    \sum_{u}
    \e^{\iu \vect{k} \cdot \vect{R}_u}
    \matr{J}_{m n}(\vect{R}_u)
\end{align}
is the Fourier transform of $\matr{J}_{m n}(\vect{R})$.

\subsection{Biquadratic interactions}
Turning to biquadratic interactions, we focus on isotropic two-site terms:
\begin{align}
    \hamil
    =
    \frac{1}{2 S^2 \hbar^4}
    \sum_{m n}
    \sum_{u v}
    I_{m n}(\vect{R}_u - \vect{R}_v)
    \qty(
        \vect{S}_{m u}
        \cdot
        \vect{S}_{n v}
    )^2
    .
\end{align}
We assume that all spins have the same spin quantum number $S$.
After a lengthy derivation, we arrived at
\begin{subequations}
\begin{align}
    A_{m n}(\vect{k})
    &=
    A_{m n}^{(1)}(\vect{k})
    +
    A_{m n}^{(2)}(\vect{k})
    +
    A_{m n}^{(3)}(\vect{k})
    ,
    \\
    A_{m n}^{(1)}(\vect{k})
    &=
    -2 S
    \tilde{I}_{m n}(\vect{k})
    \qty(
        \unitvect{z}_m
        \cdot
        \unitvect{z}_n
    )^2
    ,
    \\
    A_{m n}^{(2)}(\vect{k})
    &=
    4 S
    \tilde{I}_{m n}(\vect{k})
    \qty[
        \qty(
            \unitvect{e}_m^+
            \cdot
            \unitvect{e}_n^-
        )
        \qty(
            \unitvect{z}_m
            \cdot
            \unitvect{z}_n
        )
        +
        \qty(
            \unitvect{e}_m^+
            \cdot
            \unitvect{z}_n
        )
        \qty(
            \unitvect{e}_n^-
            \cdot
            \unitvect{z}_m
        )
    ]
    ,
    \\
    A_{m n}^{(3)}(\vect{k})
    &=
    4 S
    \kron{m}{n}
    \sum_{l = 1}^{\nbands}
    \tilde{I}_{m l}(\vect{0})
    \qty(
        \unitvect{e}_m^+
        \cdot
        \unitvect{z}_l
    )
    \qty(
        \unitvect{e}_m^-
        \cdot
        \unitvect{z}_l
    )
    ,
    \\
    B_{m n}(\vect{k})
    &=
    B_{m n}^{(1)}(\vect{k})
    +
    B_{m n}^{(2)}(\vect{k})
    ,
    \\
    B_{m n}^{(1)}(\vect{k})
    &=
    4 S
    \tilde{I}_{mn}(\vect{k})
    \qty[
        \qty(
            \unitvect{e}_m^+
            \cdot
            \unitvect{e}_n^+
        )
        \qty(
            \unitvect{z}_m
            \cdot
            \unitvect{z}_n
        )
        +
        \qty(
            \unitvect{e}_m^+
            \cdot
            \unitvect{z}_n
        )
        \qty(
            \unitvect{e}_n^+
            \cdot
            \unitvect{z}_m
        )
    ]
    ,
    \\
    B_{m n}^{(2)}(\vect{k})
    &=
    4 S
    \kron{m}{n}
    \sum_{l = 1}^{\nbands}
    \tilde{I}_{ml}(\vect{0})
    \qty(
        \unitvect{e}_m^+
        \cdot
        \unitvect{z}_l
    )^2
    .
\end{align}
\end{subequations}
using the definition
\begin{align}
    \mathcal{I}_{m n}(\vect{k})
    =
    \e^{\iu \vect{k} \cdot (\vect{r}_m - \vect{r}_n)}
    \sum_{u}
    \e^{\iu \vect{k} \cdot \vect{R}_u}
    I_{m n}(\vect{R}_u)
    .
\end{align}

\subsection{Magnon spin polarization}
For the definition of the magnon spin we compare the expectation value of the total spin operator $\vect{S}_{\text{tot}} = \sum_{i} \vect{S}_i$ in a one-magnon state $\ket{n \vect{k}} = \adj{b}_{n \vect{k}} \ket{0}$ to the classical ground-state value:
\begin{align}
    \vect{s}_{n \vect{k}}
    =
    \expval{\vect{S}_{\text{tot}}}{n \vect{k}}
    -
    \expval{\vect{S}_{\text{tot}}}{0}
    .
\end{align}
Inserting the Holstein-Primakoff transformation [\cref{eq:hpt}] for $\vect{S}_{\text{tot}}$ and keeping only terms up to quadratic order in the bosonic operators yields~\cite{okuma_magnon_2017,neumann_orbital_2020}
\begin{align}
    \vect{s}_{n \vect{k}}
    =
    -\hbar
    \sum_{m = 1}^{\nbands}
    \unitvect{z}_m
    \qty[
        \abs{T_{m n}(\vect{k})}^2
        +
        \abs{T_{m, n + \nbands}(-\vect{k})}^2
    ]
    .
\end{align}

\subsection{Linear thermal Edelstein effect}
The linear response
$
    \expval{S_\mu}
    =
    \sum_{\nu}
    \chi_{\mu \nu}
    (-\nabla_\nu T)
$
can be computed as~\cite{li_magnonic_2020}
\begin{align}
    \chi_{\mu \nu}
    &=
    \frac{1}{V T}
    \sum_{\vect{k}}
    \sum_{n = 1}^{N}
    \tau_{n \vect{k}}
    s_{n \vect{k}}^\mu
    v_{n \vect{k}}^\nu
    \varepsilon_{n \vect{k}}
    \pdv{\rho(\varepsilon_{n \vect{k}})}{\varepsilon_{n \vect{k}}}
    ,
    \label{eq:chi}
\end{align}
where $V$ is the volume of the system, $T$ is the temperature, $\tau_{n \vect{k}}$ is the relaxation time of the magnon in band $n$ with wave vector $\vect{k}$, $s_{n \vect{k}}^\mu$ is the spin expectation value, $v_{n \vect{k}}^\nu$ is the group velocity, and
$
    \rho(\varepsilon_{n \vect{k}})
    =
    \qty[
        \exp(\frac{\varepsilon_{n \vect{k}}}{k_B T})
        -
        1
    ]^{-1}
$
is the Bose function ($\kb$ is the Boltzmann constant).
We employ the constant relaxation time approximation ($\tau_{n \vect{k}} \equiv \tau$).
Note that there is no intrinsic interband contribution to $\chi_{\mu\nu}$ because it is odd under time-reversal symmetry and therefore vanishes in odd-parity-wave magnets.

\end{document}


\title{
\textit{Supplemental Material}
\vspace{0.25cm}
\hrule
\vspace{0.25cm}
Odd-Parity-Wave Magnons and Nonrelativistic Thermal Edelstein Effect
}

\author{Robin R.~Neumann\orcidlink{0000-0002-9711-3479}}
\email[Correspondence email address: ]{robin.neumann@uni-muenster.de}
\affiliation{Institute of Solid State Theory, University of Münster, D-48149 Münster, Germany}
\affiliation{Institute of Physics and Halle-Berlin-Regensburg Cluster of Excellence CCE, Martin Luther University Halle-Wittenberg, D-06099 Halle (Saale), Germany}

\author{Rodrigo Jaeschke-Ubiergo\orcidlink{0000-0002-4821-8303}}
\affiliation{Institute of Physics, Johannes Gutenberg University Mainz, D-55128 Mainz, Germany}

\author{Ricardo Zarzuela\orcidlink{0000-0003-1765-1697}}
\affiliation{Institute of Physics, Johannes Gutenberg University Mainz, D-55128 Mainz, Germany}

\author{Libor Šmejkal\orcidlink{0000-0003-1193-1372}}
\affiliation{Max Planck Institute for the Physics of Complex Systems, 01187 Dresden, Germany}
\affiliation{Max Planck Institute for Chemical Physics of Solids, 01187 Dresden, Germany}

\author{Jairo Sinova\orcidlink{0000-0002-9490-2333}}
\affiliation{Institute of Physics, Johannes Gutenberg University Mainz, D-55128 Mainz, Germany}
\affiliation{Department of Physics, Texas A\&M University, College Station, TX, USA}

\author{Alexander Mook\orcidlink{0000-0002-8599-9209}}
\affiliation{Institute of Solid State Theory, University of Münster, D-48149 Münster, Germany}

\date{\today}

\maketitle

\tableofcontents

\newpage

\section{Magnon Hamiltonian}
\subsection{Multi-component $p$-wave magnet}
Here, we present the noninteracting magnon Hamiltonian for the spin Hamiltonian
\begin{align}
    \hamil
    &=
    \frac{F}{2 S^2 \hbar^4}
    \sum_{\nns{ij}_1}
    \qty(
        \vect{S}_i
        \cdot
        \vect{S}_j
    )^2
    +
    \frac{J_3}{2 \hbar^2}
    \sum_{\nns{ij}_3}
    \vect{S}_i
    \cdot
    \vect{S}_j
    +
    \frac{J_6}{2 \hbar^2}
    \sum_{\nns{ij}_6}
    \vect{S}_i
    \cdot
    \vect{S}_j
    ,
    \label{eq:hamil_kagome}
\end{align}
where we assume the magnetic ground state of the sublattices $m$:
\begin{center}
    \begin{tabular}{c @{\hspace{1cm}} c @{\hspace{1cm}} c}
        \toprule
        $m$ & $\trp{\qty[\unitvect{e}_m^+]}$ & $\trp{\unitvect{z}}_m$
        \\
        \midrule
        1 & $\frac{1}{\sqrt{2}} \mqty(0 & 1 & \iu)$ & $\mqty(1 & 0 & 0)$
        \\
        2 & $\frac{1}{\sqrt{2}} \mqty(1 & \iu & 0)$ & $\mqty(0 & 0 & 1)$
        \\
        3 & $\frac{1}{\sqrt{2}} \mqty(\iu & 0 & 1)$ & $\mqty(0 & 1 & 0)$
        \\
        4 & $\frac{1}{\sqrt{2}} \mqty(0 & \iu & 1)$ & $\mqty(-1 & 0 & 0)$
        \\
        5 & $\frac{1}{\sqrt{2}} \mqty(\iu & 1 & 0)$ & $\mqty(0 & 0 & -1)$
        \\
        6 & $\frac{1}{\sqrt{2}} \mqty(1 & 0 & \iu)$ & $\mqty(0 & -1 & 0)$
        \\
        \bottomrule
    \end{tabular}
\end{center}

We divide the submatrices of the Hamilton matrix (see End Matter) into the bilinear contributions $\matr{A}^{\text{lin}}(\vect{k})$, $\matr{B}^{\text{lin}}(\vect{k})$ and the biquadratic ones $\matr{A}^{\text{quad}}(\vect{k})$, $\matr{B}^{\text{quad}}(\vect{k})$.
The former are
\begin{subequations}
\begin{align}
    \begin{split}
        A_{ii}^{\text{lin}}(\vect{k})
        &=
        2 S J_3
        \qty[
            1
            +
            \cos(k_x + \sqrt{3} k_y)
        ]
        +
        2 S J_6
        \qty[
            1
            +
            \cos(3 k_x - \sqrt{3} k_y)
        ]
        ,
        \qquad
        \forall i = 1, \dots, 6
        ,
    \end{split}
    \\
    \begin{split}
        B_{i, i+3}^{\text{lin}}(\vect{k})
        &=
        B_{i+3, i}^{\text{lin}}(\vect{k})
        =
        2 \iu S J_3
        \qty[
            \cos(2 k_x)
            +
            \cos(k_x - \sqrt{3} k_y)
        ]
        \\
        &\qquad
        +
        2 \iu S J_6
        \qty[
            \cos(3 k_x + \sqrt{3} k_y)
            +
            \cos(2 \sqrt{3} k_y)
        ]
        ,
        \qquad
        \forall i = 1, \dots, 3
        .
    \end{split}
\end{align}
\end{subequations}
The remaining elements are zero.
For the biquadratic interactions, the kernel reads
\begin{subequations}
\begin{align}
    \frac{\matr{A}^\text{quad}(\vect{k})}{2 S F}
    &=
    \begin{pmatrix}
        4 & c_1 & c_2 & 0 & -\iu \conj{c_1} & 0
        \\
        \conj{c_1} & 4 & c_3 & \iu c_1 & 0 & -\iu \conj{c_3}
        \\
        \conj{c_2} & \conj{c_3} & 4 & 0 & \iu c_3 & 0
        \\
        0 & -\iu \conj{c_1} & 0 & 4 & -c_1 & -c_2
        \\
        \iu c_1 & 0 & -\iu \conj{c_3} & -\conj{c_1} & 4 & -c_3
        \\
        0 & \iu c_3 & 0 & -\conj{c_2} & -\conj{c_3} & 4
    \end{pmatrix}
    ,
    \\
    \frac{\matr{B}^\text{quad}(\vect{k})}{2 S F}
    &=
    \begin{pmatrix}
        0 & c_1 & -c_2 & 0 & \iu \conj{c_1} & 0
        \\
        -\conj{c_1} & 0 & c_3 & \iu c_1 & 0 & \iu \conj{c_3}
        \\
        \conj{c_2} & -\conj{c_3} & 0 & 0 & \iu c_3 & 0
        \\
        0 & -\iu \conj{c_1} & 0 & 0 & c_1 & -c_2
        \\
        -\iu c_1 & 0 & -\iu \conj{c_3} & -\conj{c_1} & 0 & c_3
        \\
        0 & -\iu c_3 & 0 & \conj{c_2} & -\conj{c_3} & 0
    \end{pmatrix}
\end{align}
\end{subequations}
with
\begin{subequations}
\begin{align}
    c_1 &= \iu \e^{\iu k_x},
    \\
    c_2 &= -2 \iu \cos \frac{k_x + \sqrt{3} k_y}{2},
    \\
    c_3 &= \iu \e^{-\iu \frac{k_x - \sqrt{3} k_y}{2}}.
\end{align}
\end{subequations}

\subsection{Single-component $p$-wave magnet}
We consider the same Hamiltonian as before [\cref{eq:hamil_kagome}], but this time with another ground state:
\begin{center}
    \begin{tabular}{c @{\hspace{1cm}} c @{\hspace{1cm}} c}
        \toprule
        $m$ & $\trp{\qty[\unitvect{e}_m^+]}$ & $\trp{\unitvect{z}}_m$
        \\
        \midrule
        1 & $\frac{1}{\sqrt{2}} \mqty(0 & 1 & \iu)$ & $\mqty(1 & 0 & 0)$
        \\
        2 & $\frac{1}{\sqrt{2}} \mqty(1 & \iu & 0)$ & $\mqty(0 & 0 & 1)$
        \\
        3 & $\frac{1}{\sqrt{2}} \mqty(\iu & 0 & 1)$ & $\mqty(0 & 1 & 0)$
        \\
        4 & $\frac{1}{\sqrt{2}} \mqty(0 & \iu & 1)$ & $\mqty(-1 & 0 & 0)$
        \\
        5 & $\frac{1}{\sqrt{2}} \mqty(1 & \iu & 0)$ & $\mqty(0 & 0 & 1)$
        \\
        6 & $\frac{1}{\sqrt{2}} \mqty(1 & 0 & \iu)$ & $\mqty(0 & -1 & 0)$
        \\
        7 & $\frac{1}{\sqrt{2}} \mqty(0 & \iu & 1)$ & $\mqty(-1 & 0 & 0)$
        \\
        8 & $\frac{1}{\sqrt{2}} \mqty(1 & -\iu & 0)$ & $\mqty(0 & 0 & -1)$
        \\
        9 & $\frac{1}{\sqrt{2}} \mqty(1 & 0 & \iu)$ & $\mqty(0 & -1 & 0)$
        \\
        10 & $\frac{1}{\sqrt{2}} \mqty(0 & 1 & \iu)$ & $\mqty(1 & 0 & 0)$
        \\
        11 & $\frac{1}{\sqrt{2}} \mqty(1 & -\iu & 0)$ & $\mqty(0 & 0 & -1)$
        \\
        12 & $\frac{1}{\sqrt{2}} \mqty(\iu & 0 & 1)$ & $\mqty(0 & 1 & 0)$
        \\
        \bottomrule
    \end{tabular}
\end{center}
As before, we write the block matrices as a sum of the contributions from different interactions:
\begin{subequations}
\begin{align}
    \matr{A}(\vect{k})
    &=
    \matr{A}^F(\vect{k})
    +
    \matr{A}^{J_3}(\vect{k})
    +
    \matr{A}^{J_6}(\vect{k})
    ,
    \\
    \matr{B}(\vect{k})
    &=
    \matr{B}^F(\vect{k})
    +
    \matr{B}^{J_3}(\vect{k})
    +
    \matr{B}^{J_6}(\vect{k})
    ,
\end{align}
\end{subequations}
so that we have
\begin{subequations}
\begin{align}
	\frac{\matr{A}^F(\vect{k})}{2 S F}
	&=
	\begin{pmatrix}
		4 & \iu c_{1} & - \iu c_{2} & 0 & \iu \conj{c_1} & 0 & 0 & 0 & - \conj{c_2} & 0 & 0 & 0
		\\
		- \iu \conj{c_1} & 4 & \iu c_{3} & - c_{1} & 0 & 0 & 0 & 0 & 0 & 0 & 0 & \iu \conj{c_3}
		\\
		\iu \conj{c_2} & - \iu \conj{c_3} & 4 & 0 & 0 & 0 & - c_{2} & 0 & 0 & 0 & - \iu c_{3} & 0
		\\
		0 & - \conj{c_1} & 0 & 4 & - c_{1} & \iu c_{2} & 0 & 0 & 0 & 0 & 0 & - \conj{c_2}
		\\
		- \iu c_{1} & 0 & 0 & - \conj{c_1} & 4 & - c_{3} & 0 & 0 & - \conj{c_3} & 0 & 0 & 0
		\\
		0 & 0 & 0 & - \iu \conj{c_2} & - \conj{c_3} & 4 & 0 & - c_{3} & 0 & - c_{2} & 0 & 0
		\\
		0 & 0 & - \conj{c_2} & 0 & 0 & 0 & 4 & c_{1} & \iu c_{2} & 0 & \conj{c_1} & 0
		\\
		0 & 0 & 0 & 0 & 0 & - \conj{c_3} & \conj{c_1} & 4 & - c_{3} & \iu c_{1} & 0 & 0
		\\
		- c_{2} & 0 & 0 & 0 & - c_{3} & 0 & - \iu \conj{c_2} & - \conj{c_3} & 4 & 0 & 0 & 0
		\\
		0 & 0 & 0 & 0 & 0 & - \conj{c_2} & 0 & - \iu \conj{c_1} & 0 & 4 & - \iu c_{1} & - \iu c_{2}
		\\
		0 & 0 & \iu \conj{c_3} & 0 & 0 & 0 & c_{1} & 0 & 0 & \iu \conj{c_1} & 4 & \iu c_{3}
		\\
		0 & - \iu c_{3} & 0 & - c_{2} & 0 & 0 & 0 & 0 & 0 & \iu \conj{c_2} & - \iu \conj{c_3} & 4
	\end{pmatrix}
    ,
    \\
    \frac{\matr{B}^F(\vect{k})}{2 S F}
    &=
    \begin{pmatrix}
        0 & \iu c_{1} & \iu c_{2} & 0 & \iu \conj{c_1} & 0 & 0 & 0 & - \conj{c_2} & 0 & 0 & 0
        \\
        \iu \conj{c_1} & 0 & \iu c_{3} & - c_{1} & 0 & 0 & 0 & 0 & 0 & 0 & 0 & \iu \conj{c_3}
        \\
        \iu \conj{c_2} & \iu \conj{c_3} & 0 & 0 & 0 & 0 & c_{2} & 0 & 0 & 0 & \iu c_{3} & 0
        \\
        0 & - \conj{c_1} & 0 & 0 & - c_{1} & \iu c_{2} & 0 & 0 & 0 & 0 & 0 & \conj{c_2}
        \\
        \iu c_{1} & 0 & 0 & - \conj{c_1} & 0 & c_{3} & 0 & 0 & \conj{c_3} & 0 & 0 & 0
        \\
        0 & 0 & 0 & \iu \conj{c_2} & \conj{c_3} & 0 & 0 & c_{3} & 0 & - c_{2} & 0 & 0
        \\
        0 & 0 & \conj{c_2} & 0 & 0 & 0 & 0 & c_{1} & \iu c_{2} & 0 & \conj{c_1} & 0
        \\
        0 & 0 & 0 & 0 & 0 & \conj{c_3} & \conj{c_1} & 0 & c_{3} & - \iu c_{1} & 0 & 0
        \\
        - c_{2} & 0 & 0 & 0 & c_{3} & 0 & \iu \conj{c_2} & \conj{c_3} & 0 & 0 & 0 & 0
        \\
        0 & 0 & 0 & 0 & 0 & - \conj{c_2} & 0 & - \iu \conj{c_1} & 0 & 0 & - \iu c_{1} & \iu c_{2}
        \\
        0 & 0 & \iu \conj{c_3} & 0 & 0 & 0 & c_{1} & 0 & 0 & - \iu \conj{c_1} & 0 & \iu c_{3}
        \\
        0 & \iu c_{3} & 0 & c_{2} & 0 & 0 & 0 & 0 & 0 & \iu \conj{c_2} & \iu \conj{c_3} & 0
    \end{pmatrix}
\end{align}
\end{subequations}
for the biquadratic interaction,
\begin{subequations}
\begin{align}
    \frac{
        \matr{A}^{J_3}(\vect{k})
    }{
        2 J_3 S
    }
    &=
    \begin{pmatrix}
        1 & 0 & 0 & 0 & 0 & 0 & 0 & 0 & 0 & c_{6} & 0 & 0
        \\
        0 & 1 & 0 & 0 & c_{4} & 0 & 0 & 0 & 0 & 0 & 0 & 0
        \\
        0 & 0 & 1 & 0 & 0 & 0 & 0 & 0 & 0 & 0 & 0 & c_{6}
        \\
        0 & 0 & 0 & 1 & 0 & 0 & c_{6} & 0 & 0 & 0 & 0 & 0
        \\
        0 & c_{4} & 0 & 0 & 1 & 0 & 0 & 0 & 0 & 0 & 0 & 0
        \\
        0 & 0 & 0 & 0 & 0 & 1 & 0 & 0 & c_{6} & 0 & 0 & 0
        \\
        0 & 0 & 0 & c_{6} & 0 & 0 & 1 & 0 & 0 & 0 & 0 & 0
        \\
        0 & 0 & 0 & 0 & 0 & 0 & 0 & 1 & 0 & 0 & c_{4} & 0
        \\
        0 & 0 & 0 & 0 & 0 & c_{6} & 0 & 0 & 1 & 0 & 0 & 0
        \\
        c_{6} & 0 & 0 & 0 & 0 & 0 & 0 & 0 & 0 & 1 & 0 & 0
        \\
        0 & 0 & 0 & 0 & 0 & 0 & 0 & c_{4} & 0 & 0 & 1 & 0
        \\
        0 & 0 & c_{6} & 0 & 0 & 0 & 0 & 0 & 0 & 0 & 0 & 1
    \end{pmatrix}
    ,
    \\
    \frac{
        \matr{B}^{J_3}(\vect{k})
    }{
        2 J_3 S
    }
    &=
    \begin{pmatrix}
        0 & 0 & 0 & \iu c_{4} & 0 & 0 & \iu c_{5} & 0 & 0 & 0 & 0 & 0
        \\
        0 & 0 & 0 & 0 & 0 & 0 & 0 & c_{5} & 0 & 0 & c_{6} & 0
        \\
        0 & 0 & 0 & 0 & 0 & \iu c_{4} & 0 & 0 & \iu c_{5} & 0 & 0 & 0
        \\
        \iu c_{4} & 0 & 0 & 0 & 0 & 0 & 0 & 0 & 0 & \iu c_{5} & 0 & 0
        \\
        0 & 0 & 0 & 0 & 0 & 0 & 0 & c_{6} & 0 & 0 & c_{5} & 0
        \\
        0 & 0 & \iu c_{4} & 0 & 0 & 0 & 0 & 0 & 0 & 0 & 0 & \iu c_{5}
        \\
        \iu c_{5} & 0 & 0 & 0 & 0 & 0 & 0 & 0 & 0 & \iu c_{4} & 0 & 0
        \\
        0 & c_{5} & 0 & 0 & c_{6} & 0 & 0 & 0 & 0 & 0 & 0 & 0
        \\
        0 & 0 & \iu c_{5} & 0 & 0 & 0 & 0 & 0 & 0 & 0 & 0 & \iu c_{4}
        \\
        0 & 0 & 0 & \iu c_{5} & 0 & 0 & \iu c_{4} & 0 & 0 & 0 & 0 & 0
        \\
        0 & c_{6} & 0 & 0 & c_{5} & 0 & 0 & 0 & 0 & 0 & 0 & 0
        \\
        0 & 0 & 0 & 0 & 0 & \iu c_{5} & 0 & 0 & \iu c_{4} & 0 & 0 & 0
    \end{pmatrix}
    ,
\end{align}
\end{subequations}
for the bilinear $J_3$ interaction, and
\begin{subequations}
\begin{align}
	\begin{split}
        \frac{
		      \matr{A}^{J_6}(\vect{k})
        }{
		      2 J_6 S
        }
		&=
		\begin{pmatrix}
			1 & 0 & 0 & 0 & 0 & 0 & 0 & 0 & 0 & c_{7} & 0 & 0
			\\
			0 & 1 & 0 & 0 & c_{8} & 0 & 0 & 0 & 0 & 0 & 0 & 0
			\\
			0 & 0 & 1 & 0 & 0 & 0 & 0 & 0 & 0 & 0 & 0 & c_{7}
			\\
			0 & 0 & 0 & 1 & 0 & 0 & c_{7} & 0 & 0 & 0 & 0 & 0
			\\
			0 & c_{8} & 0 & 0 & 1 & 0 & 0 & 0 & 0 & 0 & 0 & 0
			\\
			0 & 0 & 0 & 0 & 0 & 1 & 0 & 0 & c_{7} & 0 & 0 & 0
			\\
			0 & 0 & 0 & c_{7} & 0 & 0 & 1 & 0 & 0 & 0 & 0 & 0
			\\
			0 & 0 & 0 & 0 & 0 & 0 & 0 & 1 & 0 & 0 & c_{8} & 0
			\\
			0 & 0 & 0 & 0 & 0 & c_{7} & 0 & 0 & 1 & 0 & 0 & 0
			\\
			c_{7} & 0 & 0 & 0 & 0 & 0 & 0 & 0 & 0 & 1 & 0 & 0
			\\
			0 & 0 & 0 & 0 & 0 & 0 & 0 & c_{8} & 0 & 0 & 1 & 0
			\\
			0 & 0 & c_{7} & 0 & 0 & 0 & 0 & 0 & 0 & 0 & 0 & 1
		\end{pmatrix}
		,
	\end{split}
	\\
	\begin{split}
        \frac{
		      \matr{B}^{J_6}(\vect{k})
        }{
		      2 J_6 S
        }
		&=
		\begin{pmatrix}
			0 & 0 & 0 & \iu c_{8} & 0 & 0 & \iu c_{9} & 0 & 0 & 0 & 0 & 0
			\\
			0 & 0 & 0 & 0 & 0 & 0 & 0 & c_{9} & 0 & 0 & c_{7} & 0
			\\
			0 & 0 & 0 & 0 & 0 & \iu c_{8} & 0 & 0 & \iu c_{9} & 0 & 0 & 0
			\\
			\iu c_{8} & 0 & 0 & 0 & 0 & 0 & 0 & 0 & 0 & \iu c_{9} & 0 & 0
			\\
			0 & 0 & 0 & 0 & 0 & 0 & 0 & c_{7} & 0 & 0 & c_{9} & 0
			\\
			0 & 0 & \iu c_{8} & 0 & 0 & 0 & 0 & 0 & 0 & 0 & 0 & \iu c_{9}
			\\
			\iu c_{9} & 0 & 0 & 0 & 0 & 0 & 0 & 0 & 0 & \iu c_{8} & 0 & 0
			\\
			0 & c_{9} & 0 & 0 & c_{7} & 0 & 0 & 0 & 0 & 0 & 0 & 0
			\\
			0 & 0 & \iu c_{9} & 0 & 0 & 0 & 0 & 0 & 0 & 0 & 0 & \iu c_{8}
			\\
			0 & 0 & 0 & \iu c_{9} & 0 & 0 & \iu c_{8} & 0 & 0 & 0 & 0 & 0
			\\
			0 & c_{7} & 0 & 0 & c_{9} & 0 & 0 & 0 & 0 & 0 & 0 & 0
			\\
			0 & 0 & 0 & 0 & 0 & \iu c_{9} & 0 & 0 & \iu c_{8} & 0 & 0 & 0
		\end{pmatrix}
	\end{split}
\end{align}
\end{subequations}
for the bilinear $J_6$ coupling.
We have defined the auxiliary complex-valued functions
\begin{subequations}
\begin{align}
	\begin{split}
		c_{1}
		&=
		\e^{\iu k_{x}}
		,
	\end{split}
	\\
	\begin{split}
		c_{2}
		&=
		\e^{\iu \frac{k_x + \sqrt{3} k_y}{2}}
		,
	\end{split}
	\\
	\begin{split}
		c_{3}
		&=
		\e^{-\iu \frac{k_x - \sqrt{3} k_y}{2}}
		,
	\end{split}
    \\
	\begin{split}
		c_{4}
		&=
		\cos(2 k_x)
		,
	\end{split}
	,
	\\
	\begin{split}
		c_{5}
		&=
		\cos(k_x + \sqrt{3} k_y)
		,
	\end{split}
	\\
	\begin{split}
		c_{6}
		&=
		\cos(k_x - \sqrt{3} k_y)
		,
	\end{split}
    \\
	\begin{split}
		c_{7}
		&=
		\cos(3 k_x + \sqrt{3} k_y)
		,
	\end{split}
	\\
	\begin{split}
		c_{8}
		&=
		\cos(2 \sqrt{3} k_y)
		,
	\end{split}
	\\
	\begin{split}
		c_{9}
		&=
		\cos(3 k_x - \sqrt{3} k_y)
		.
	\end{split}
\end{align}
\end{subequations}

\subsection{Single-component $f$-wave magnet}
For the triangular-lattice model, the Hamiltonian has been defined as
\begin{align}
    \hamil
    &=
    \frac{F}{2 S^2 \hbar^4}
    \sum_{\langle ij \rangle_1}
    \qty(
        \vect{S}_i
        \cdot
        \vect{S}_j
    )^2
    +
    \frac{J_2}{2 \hbar^2}
    \sum_{\langle ij \rangle_2}
    \vect{S}_i
    \cdot
    \vect{S}_j
    +
    \frac{J_\perp}{2 \hbar^2}
    \sum_{\langle ij \rangle_\perp}
    \vect{S}_i
    \cdot
    \vect{S}_j
    ,
\end{align}
and the ground state under consideration is
\begin{center}
    \begin{tabular}{c @{\hspace{1cm}} c @{\hspace{1cm}} c}
        \toprule
        $m$ & $\trp{\qty[\unitvect{e}_m^+]}$ & $\trp{\unitvect{z}}_m$
        \\
        \midrule
        1 & $\frac{1}{\sqrt{2}} \mqty(0 & 1 & \iu)$ & $\mqty(1 & 0 & 0)$
        \\
        2 & $\frac{1}{\sqrt{2}} \mqty(1 & \iu & 0)$ & $\mqty(0 & 0 & 1)$
        \\
        3 & $\frac{1}{\sqrt{2}} \mqty(\iu & 0 & 1)$ & $\mqty(0 & 1 & 0)$
        \\
        4 & $\frac{1}{\sqrt{2}} \mqty(0 & \iu & 1)$ & $\mqty(-1 & 0 & 0)$
        \\
        5 & $\frac{1}{\sqrt{2}} \mqty(1 & -\iu & 0)$ & $\mqty(0 & 0 & -1)$
        \\
        6 & $\frac{1}{\sqrt{2}} \mqty(1 & 0 & \iu)$ & $\mqty(0 & -1 & 0)$
        \\
        \bottomrule
    \end{tabular}
\end{center}
We write
\begin{subequations}
\begin{align}
    \matr{A}(\vect{k})
    &=
    \matr{A}^F(\vect{k})
    +
    \matr{A}^{J_2}(\vect{k})
    +
    \matr{A}^{J_\perp}(\vect{k})
    ,
    \\
    \matr{B}(\vect{k})
    &=
    \matr{B}^F(\vect{k})
    +
    \matr{B}^{J_2}(\vect{k})
    +
    \matr{B}^{J_\perp}(\vect{k})
    ,
\end{align}
\end{subequations}
and obtain
\begin{subequations}
\begin{align}
    \begin{split}
        \frac{
            \matr{A}^{F}(\vect{k})
        }{
            2 S^3 F
        }
        &=
        \begin{pmatrix}
            6 & \iu c_{1} + \iu \conj{c_2} + \iu c_{3} & - \iu \conj{c_1} - \iu c_{2} - \iu \conj{c_3} & 0 & 0 & 0
            \\
            - \iu \conj{c_1} - \iu c_{2} - \iu \conj{c_3} & 6 & \iu c_{1} + \iu \conj{c_2} + \iu c_{3} & 0 & 0 & 0
            \\
            \iu c_{1} + \iu \conj{c_2} + \iu c_{3} & - \iu \conj{c_1} - \iu c_{2} - \iu \conj{c_3} & 6 & 0 & 0 & 0
            \\
            0 & 0 & 0 & 6 & c_{1} + \conj{c_2} + c_{3} & \iu \conj{c_1} + \iu c_{2} + \iu \conj{c_3}
            \\
            0 & 0 & 0 & \conj{c_1} + c_{2} + \conj{c_3} & 6 & - c_{1} - \conj{c_2} - c_{3}
            \\
            0 & 0 & 0 & - \iu c_{1} - \iu \conj{c_2} - \iu c_{3} & - \conj{c_1} - c_{2} - \conj{c_3} & 6
        \end{pmatrix}
    \end{split}
    \\
    \begin{split}
        \frac{
            \matr{B}^{F}(\vect{k})
        }{
            2 S^3 F
        }
        &=
        \begin{pmatrix}
            0 & \iu c_{1} + \iu \conj{c_2} + \iu c_{3} & \iu \conj{c_1} + \iu c_{2} + \iu \conj{c_3} & 0 & 0 & 0
            \\
            \iu \conj{c_1} + \iu c_{2} + \iu \conj{c_3} & 0 & \iu c_{1} + \iu \conj{c_2} + \iu c_{3} & 0 & 0 & 0
            \\
            \iu c_{1} + \iu \conj{c_2} + \iu c_{3} & \iu \conj{c_1} + \iu c_{2} + \iu \conj{c_3} & 0 & 0 & 0 & 0
            \\
            0 & 0 & 0 & 0 & c_{1} + \conj{c_2} + c_{3} & \iu \conj{c_1} + \iu c_{2} + \iu \conj{c_3}
            \\
            0 & 0 & 0 & \conj{c_1} + c_{2} + \conj{c_3} & 0 & c_{1} + \conj{c_2} + c_{3}
            \\
            0 & 0 & 0 & \iu c_{1} + \iu \conj{c_2} + \iu c_{3} & \conj{c_1} + c_{2} + \conj{c_3} & 0
        \end{pmatrix}
    \end{split}
    \\
    \begin{split}
        \matr{A}^{J_2}(\vect{k})
        &=
        2 S J_2
        (c_{4} + c_{5} + c_{6} - 3)
        \idMatr
        ,
    \end{split}
    \\
    \begin{split}
        \matr{B}^{J_2}(\vect{k})
        &=
        \matr{0}
        ,
    \end{split}
    \\
    \begin{split}
        \matr{A}^{\perp}(\vect{k})
        &=
        2 S J_\perp
        \idMatr
        ,
    \end{split}
    \\
    \begin{split}
        \matr{B}^{\perp}(\vect{k})
        &=
        2 S J_\perp
        \begin{pmatrix}
            0 & 0 & 0 & \iu c_{7} & 0 & 0
            \\
            0 & 0 & 0 & 0 & c_{7} & 0
            \\
            0 & 0 & 0 & 0 & 0 & \iu c_{7}
            \\
            \iu c_{7} & 0 & 0 & 0 & 0 & 0
            \\
            0 & c_{7} & 0 & 0 & 0 & 0
            \\
            0 & 0 & \iu c_{7} & 0 & 0 & 0
        \end{pmatrix}
        ,
    \end{split}
\end{align}
\end{subequations}
where we have introduced
\begin{subequations}
\begin{align}
    \begin{split}
        c_1 &= \exp(\iu k_x)
    \end{split}
    \\
    \begin{split}
        c_2 &= \exp(\iu \frac{k_x + \sqrt{3} k_y}{2})
    \end{split}
    \\
    \begin{split}
        c_3 &= \exp(-\iu \frac{k_x - \sqrt{3} k_y}{2})
    \end{split}
    \\
    \begin{split}
        c_4 &= \cos(\frac{3 k_x + \sqrt{3} k_y}{2})
        ,
    \end{split}
    \\
    \begin{split}
        c_5 &= \cos(\sqrt{3} k_y)
        ,
    \end{split}
    \\
    \begin{split}
        c_6 &= \cos(\frac{3 k_x - \sqrt{3} k_y}{2})
        ,
    \end{split}
    \\
    \begin{split}
        c_7 &= \cos k_z
        .
    \end{split}
\end{align}
\end{subequations}

\section{Spin symmetry analysis}
In this section we describe the structure of the spin point groups of the three models of odd-parity-wave magnets presented in this work. We obtain the elements of the spin space group numerically \cite{Shinohara2024}, and then we write the spin point group in the following way

\begin{equation}
\mathcal{R}_s = \mathbf{r}_{\rm so} \times \mathbf{G}_{\rm st} \times \mathbf{R}_{\rm s} \text{ ,}
\end{equation}
where $\mathbf{r}_{\rm so}$ is the spin-only group, $\mathbf{G}_{\rm st}$ captures the point group part of the nontrivial spin translation group \cite{Watanabe2024}, in which all translational elements are projected into the identity, and $\mathbf{R}_{\rm s}$ is a spin point group of generally distinct elements acting in real and spin space. In the remainder of this section, we use the term \textit{spin translation group} as shorthand for the point part of the nontrivial spin translation group \cite{Watanabe2024}.

Since all models in this work exhibit a noncoplanar spin arrangement, the spin-only group only contains the identity, and does not play a role in the label of the group.
 We recognize and factorize out the spin-translation group $\mathbf{G}_{\rm st}$, whose elements have the form $[R_{\rm s}||E]$ or $[R_{\rm s} \mathcal{T}||\mathcal{T}]$, inherited form the spin space group elements $[R_{\rm s}||E|\boldsymbol{\tau}]$ and $[R_{\rm s}\mathcal{T}||\mathcal{T}|\boldsymbol{\tau}]$, respectively.
 Here, $R_{\rm s}$ is a proper rotation acting in space, $E$ is the identity, $\mathcal{T}$ denotes time-reversal symmetry, and $\boldsymbol{\tau}$ is a translation. 
 
 The remaining part of the group, in which the identity in real space appears only paired with the identity in spin space $[E||E]$, is identified with one of the tabulated spin point groups in Ref. \cite{Litvin1977}. In this construction, one builds a spin point group $\mathbf{R}_s$ by using a spin parent group $\mathbf{B}$, a crystallographic parent group $\mathbf{G}$, and an invariant (normal) subgroup  $\mathbf{r} \trianglelefteq \mathbf{G}$. The spin parent group is isomorphic to the factor group $\mathbf{G}/\mathbf{r}$, in such a way that each element of $\mathbf{B}$ is paired with one element of the coset decomposition of $\mathbf{G}$ with respect to $\mathbf{r}$.

\subsection{Multi-component $p$-wave magnet}
The model of Fig.~2 of the main text has the spin point group ${}^{\bar{1}}1 {}^{2_x}m {}^{2_y}m {}^{1}m$.
Here, $\mathbf{G}_{\rm st}={}^{\bar{1}}1$ is the spin translation group coming from the spin translation element $[\mathcal{T}||\mathcal{T}|\boldsymbol{\tau}_1]$, and ${}^{2_x}m {}^{2_y}m {}^{1}m$ corresponds to the spin point group that one builds with a spin parent group $\mathbf{B}=222$, crystallographic real-space parent group $\mathbf{G}=mmm$, and invariant subgroup $\mathbf{r}=m$ (spin point group No.~74 in Ref.~\cite{Litvin1977}).

In our setting, the generators of this group are $[\mathcal{T}||\mathcal{T}]$, $[C_{2,xy}||m_{u}]$, $[C_{2,x\bar{y}}||m_{v}]$ and $[E||m_z]$, with $C_{2,xy}$, and $C_{2,x\bar{y}}$ two-fold spin rotations around the diagonal axes $(1,1,0)$ and $(1,-1,0)$ (in Cartesian coordinates). $m_u$, and $m_v$ are mirror planes normal to the directions $k_x$ and $k_y$ indicated in Fig.~2(a) in the main text, that are rotated by \SI{-30}{\degree} with respect to $x$, and $y$ axes, and $m_z$ is a mirror plane normal to the $z$ axis.

\subsection{Single-component $p$-wave magnet}
The model presented in Fig.~3 of the main text has the spin point group ${}^{2_z}1 {}^{\bar{1}}1 {}^{2_y}2/{}^{1}m$.
Here, the first symbol ${}^{2_z}1 {}^{\bar{1}}1$ represents $\mathbf{G}_{\rm st}=\{[E||E], [C_{2z}||E], [C_{2z}\mathcal{T}||\mathcal{T}], [\mathcal{T}||\mathcal{T}|] \}$, with $C_{2z}$ a two-fold spin rotation about the $z$ axis.
The remaining part $\mathbf{R}_{\rm s}={}^{2_y}2/{}^{1}m$ is the spin point group associated with spin parent group $\mathbf{B}= 2$, real space parent group $\mathbf{G}=2/m$ and invariant subgroup $\mathbf{r}=m$ (No. 18 in Ref. \cite{Litvin1977}).

In our setting, the specific form of the generators appearing in the group name is: $[C_{2z}|| E]$, $[ \mathcal{T}|| \mathcal{T}]$, $[ C_{2y}|| C_{2z}]$ and $[E ||m_z]$.
 The first of the generators $[C_{2z}|| E]$ constrains the spin polarization in momentum space to be collinear and out-of-plane, $[ \mathcal{T}|| \mathcal{T}]$ enforces antisymmetric spin polarization, and the element $[ C_{2y}|| C_{2z}]$  imposes a single spin unpolarized line, that crosses the $\Gamma$ point, but whose general position is not pinned to special symmetry line and is generally curved.

\subsection{Single-component $f$-wave magnet}
The model presented in Fig.~4 of the main text has spin point group $^{\bar{3}} 1 ^m 6 / ^1 m ^m m ^1 m$.
Here, the spin translation group is $\mathbf{G}_{\rm st}= {}^{\bar{3}}1=\{[E||E], [ C_{3[111]}\mathcal{T}||\mathcal{T}],[ C_{3[111]}^{-1}|| E],[ \mathcal{T}||\mathcal{T}],[ C_{3[111]}|| E],[ C_{3[111]}^{-1}\mathcal{T}|| \mathcal{T}] \}$, where $C_{3[111]}$ is a three-fold rotation in spin space along the axis $(1, 1, 1)$ (in Cartesian coordinates).
The remaining part of the group $\mathbf{R}_{\rm s} =  {}^m 6 / {}^1 m {}^m m {}^1 m$ can be built with the spin parent group $\mathbf{B} = m$, the real-space parent group $\mathbf{G}=6/mmm$ and invariant subgroup $\mathbf{r}=\bar{6}m2$ (No.~445 in Ref.~\cite{Litvin1977}).

In our setting, the complete spin point group $^{\bar{3}} 1 ^m 6 / ^1 m ^m m ^1 m$ is generated by the spin symmetries: $[C_{3[111]}\mathcal{T}|| \mathcal{T}]$, $[C_{2[1\bar{1}0]}\mathcal{T} || C_{6z} \mathcal{T}]$, $[ E|| m_z]$, $[ C_{2[1\bar{1}0]}\mathcal{T}|| m_x \mathcal{T}]$ and $[E||m_y]$. Here, each element corresponds to one of the symbols in the name $^{\bar{3}} 1 ^m 6 / ^1 m ^m m ^1 m$. $C_{2[1\bar{1}0]}$ is a two-fold spin rotation along the axis $(1, -1, 0)$ (in Cartesian coordinates).

The $f$-wave character can be extracted from the structure of this group. The spin translation group $\mathbf{G}_{\rm st}$ imposes a collinear spin polarization along the $[111]$ direction. Note that $C_{2[1\bar{1}0]}$ reverses a vector along $[111]$. It follows that those spin symmetries with $C_{2[1\bar{1}0]}$ in the spin part will connect regions in the Brillouin zone with opposite spin polarization. From $[C_{2[1\bar{1}0]}||C_{6z}]$ and $[C_{2[1\bar{1}0]}||m_x]$ one concludes the presence of 3 spin-unpolarized nodal planes that intersect the K and K' points. 

\section{Thermal linear Edelstein effect}
\begin{figure}
    \centering
    \includegraphics[width=.5\linewidth]{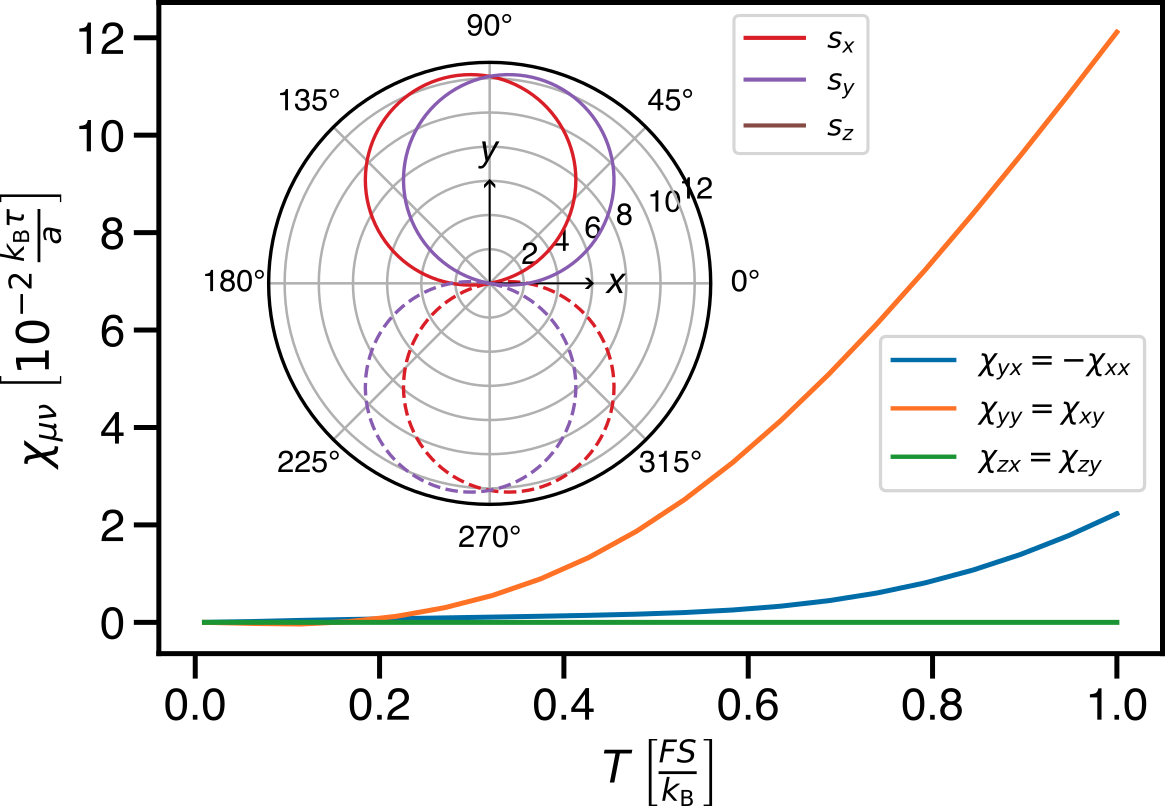}
    \caption{%
        Temperature-dependent linear thermal Edelstein effect of the multi-component $p$-wave magnet with parameters $J_3 = 2 J_6 = F$.
        Here, $\tau$ is the relaxation time and $a$ is the nearest-neighbor distance.
        The real-space index $\nu$ refers to the coordinate system that has been aligned with the $k_x$ and $k_y$ axes in Fig.~2(a) of the main text.
        Inset: Nonequilibrium spin polarization as a function of the direction of the (negative) temperature gradient $-\grad T$ at $T = F S / \kb$.
        The angle of \SI{0}{\degree} corresponds to $-\grad T \parallel \unitvect{x}$.
        Different colors correspond to different spin components.
        Solid and dashed lines indicate positive and negative sign, respectively.%
    }
    \label{fig:edelstein_noncoplanar_kagome}
\end{figure}
As for the single-component $p$-wave magnet, we have computed the linear-response coefficients $\chi_{\mu\nu}$ within the constant-relaxation time approximation ($\tau_{n \vect{k}} \equiv \tau$) for the multi-component $p$-wave magnet (\cref{fig:edelstein_noncoplanar_kagome}).
Because of the spin point group there are only two nonzero independent components ($\chi_{yx} = -\chi_{xx}$ and $\chi_{yy} = \chi_{xy}$), which are activated by temperature.
Here, the real-space coordinate system, upon which the second index $\nu$ depends, is aligned with the $k_x$ and $k_y$ axes in Fig.~2(a) of the main text.
In the inset of \cref{fig:edelstein_noncoplanar_kagome}, we present the dependence of the nonequilibrium spin polarization on the direction of the applied temperature gradient.
For the $x$ and $y$ spin components, we respectively find one lobe with positive (solid line) and one lobe with negative sign (dashed line) reminiscent of a $p$ orbital.
The two $p$-orbital-like patterns are mostly extended along the $y$ direction implying a prominent anisotropy with large responses for temperature gradients applied in the direction of the lobes.
The elements with $\mu = z$ vanish because the magnon bands are unpolarized with respect to $s_z$ rendering the corresponding Edelstein response zero.
This situation is reversed compared to the single-component $p$-wave magnet, where only the $\mu = z$ component survives, as shown in the main text.

\bibliography{refs}